\documentstyle[preprint,aps]{revtex}

\begin{document}
\tolerance 50000

\draft

\title{Real-Space Renormalization Group Methods Applied to Quantum Lattice 
Hamiltonians}
\author{Miguel A. Mart\'{\i}n-Delgado$^{1}$ } 
\address{ $^{1}$Departamento de
F\'{\i}sica Te\'orica I, Universidad Complutense.  28040-Madrid, Spain
 \\
}

\date{October 96}
\maketitle
\widetext

\vspace*{-1.0truecm}

\begin{abstract}
\begin{center}
\parbox{14cm}{ I review recent work and some new results, 
performed in collaboration with G. Sierra, on the 
Real-Space Renormalization group method applied to quantum spin lattice systems
mainly in spatial dimensions one and two, and to spin ladders which are somehow
in between. The first part of these notes is devoted to non-interacting systems in
1D and 2D and the role played by the correlations between blocks.
The second part comprises interacting systems in 1D, spin ladders and 2D using the
standard BRG method.
}
\end{center}
\end{abstract}

\vspace{2cm} {\em Proceedings of the El Escorial Summer School 1996 on 
STRONGLY CORRELATED MAGNETIC AND SUPERCONDUCTING SYSTEMS}

\section{Introduction: Brief History of Real-Space RG Methods}

The Real-Space Renormalization Group Method has undergone a great revival
since the Densty Matrix RG was introduce by White in 1992 \cite{white}, \cite{whitenote}.
Nowdays, this method is considered a powerful numerical tool to get non-perturbative
results, specially for interacting systems in 1D although some recent advances
for 2D systems have been obtained \cite{white2D}, \cite{whitenote}.
Despite being numerical, the DMRG has also been a source of inspiration
for analytical studies and this is the framework of the present notes.
Moreover, much of the El Escorial Summer School has been devoted to 
real-space RG methods \cite{whitenote}, \cite{nishinonote}, \cite{pereznote}.

The Renormalization Group method has become one of the basic concepts 
in Physics, ranging from areas such as Quantum Field Theory and 
Statistical Mechanics to Condensed Matter Physics. The many interesting
and relevant models encountered in these fields are usually not
exactly  solvable except for some privileged cases in one dimension.
It is then when we resort to the RG method to retrieve the essential
features of those systems in order to have a qualitative understanding
of what the physics of the model is all about. This understanding is
usually recasted in the  form of a RG-flow diagram were the different
possible behaviours of  the model leap to the eyes.

Many authors in the past have contributed significantly to the idea of
renormalization and it is out of the scope of these notes to give a
detailed account on this issue. 

We shall be dealing with the the
version of the RG as introduced by Wilson
 \cite{wilson} and Anderson \cite{anderson1}
 in their treatment of the Kondo problem, and subsequent  developments
of these ideas carried out by Drell et al. at the SLAC group 
\cite{drell} and Pfeuty et al. \cite{jullien}.

Real space Renormalization Group (RG) methods originated
from the study of the Kondo problem by Wilson
\cite{wilson}.
 It was clear from the beginning that one could not hope
to achieve the accuracy  Wilson obtained for the Kondo
problem when dealing with more complicated many-body 
quantum Hamiltonians. The key difference is  that in the
Kondo model there exists  a {\em recursion relation} for
Hamiltonians at each step  of the RG-elimination of
degrees of freedom.  The existence of such recursion relation
facilitates enormously the work, but as it  happens it is
specific of {\em impurity problems}.

From the numerical point of view, the Block
Renormalization Group (BRG)
 procedure proved to be not  fully reliable in the past
particularly in comparison with other numerical
approaches, such as  the Quantum MonteCarlo method which
were being developed at the same time. This was one of 
the reasons why the BRG methods remained undeveloped
during the '80's until the begining of  the '90's when
they are making a comeback as one of the most powerful
numerical tools  when dealing with zero temperature
properties of many-body systems, a situation where the 
Quantum MonteCarlo methods happen to be particularly badly
behaved as far as fermionic  systems is concerned
\cite{hirsch}.

As it happens,  the BRG gives a good qualitative picture
of many properties  exhibited by quantum lattice
Hamiltonians: Fixed points, RG-flow, phases of the system
etc. as well as good quantitative results for some
properties such as ground state energy and others 
\cite{drell}, \cite{jullien},  \cite{hirschII},
\cite{rabin}. 
 However in some important instances the BRG method is off
the correct values of critical exponents by a  sensible
amount.

 The origin of the density matrix RG method relies on the special
treatment carried out  by White and Noack \cite{white-noack} on the
1D tight-binding model,  the lattice version of a single particle in
a box. It was Wilson \cite{wilson-unp} 
the first to point out the relevance  of this
simple model in understanding the sometimes bad numerical performance
of the  standard Block Renormalization Group (BRG) method. In
reference \cite{white-noack} the  authors proposed a method called
Combination of Boundary Conditions (CBC) which  performs extremely
well as compared to the exact known solution of the model.  Recently,
we have studied the role played by the boundary conditions in the
real-space  renormalization group method \cite{bc-germanyo} by
constructing a new analytical BRG-method which is able to give the
exact ground state  of the model and the correct $1/N^2$-law for the
energy of the first excited state in the  large $N$(size)-limit.

\noindent Yet another branch of applications of DMRG inspired ideas
is to use the Superblock formalism \cite{white-noack} without 
resorting to a Density Matrix. For instance, in \cite{kl} it has
been found that the application of this formalism to
the anisotropic Heisenberg model in 1D successfully improves
the standard BRG results of Rabin \cite{rabin}.

\noindent The Density Matrix RG has been originally devised to deal
with quantum lattice Hamiltonians. However, recent new applications
have been developed by Nishino and coworkers \cite{nishinonote},
\cite{nishino1}, \cite{nishino-okunishi}
to address the renormalization of classical lattice models. 
One of the outcomes of these studies has been to state the
relationship between Baxter's corner transfer matrix
formalism and the Density Matrix RG of White
(in 1D).

The number of new developments on this subject is constantly growing
and it is not possible to give a full account of all of them here.
More applications can be found in the rest of contributions to these
proceedings devoted to real-space RG methods.


\section{Review of Standard 
Block Renormalization Group Methods (BRG)}

For the sake of completeness and to set up the notation used throughout these notes,
 the block renormalization group method is revisited in this section along the lines of
 a new  and unified reformulation of it based on the
idea of the {\em intertwiner operator} $T$
 to be discussed  below. This treatment is by all means equivalent to the standard
approach presented by S.R. White in his contribution to this volume.
This formulation has recently allowed us to introduce
the new Variational and Fokker-Planck  DMRG methods \cite{dm-germanyo}
 on equal footing
as the standard BRG method. For a more extensive account on this
method we refer to \cite{jullienlibro}  and chapter 11 of reference
\cite{jaitisi} and references therein.

Let us first summarize the main features of the real-space RG. The
problem that one faces   generically is that of diagonalizing a
quantum lattice Hamiltonian $H$, i.e.,

\begin{equation} H |\psi > = E |\psi >                         \label{1}
 \end{equation}

\noindent  where $|\psi >$ is a state in the Hilbert space $\cal H$.
If the lattice has N sites  and there are $k$ possible states per
site then the dimension of $\cal H$ is simply 

\begin{equation} dim{ \cal H} = k^N                             
\label{2}
 \end{equation}

\noindent As a matter of illustration we cite the following
examples:  $k=4$  (Hubbard model), $k=3$ (t-J model), $k=2$
(Heisenberg model) etc.

\noindent When $N$ is large enough the eigenvalue problem (\ref{1})
is out of the capability of  any human or computer means unless the
model turns out to be integrable which only happens in some instances
in $d=1$.

\noindent  These facts open the door to a variety of approximate
methods among which the RG-approach,  specially when combined with
other techniques (e.g. numerical, variational etc.), is one of the 
most relevant. The main idea of the RG-method is the mode elimination
or thinning of the degrees  of freedom followed by an iteration which
reduces the number of variables step by step until  a more manageable
situation is reached. These intuitive ideas give rise to a well
defined  mathematical description of the RG-approach to the low lying
spectrum of quantum lattice  hamiltonians.

To carry out the RG-program it will be useful to introduce the
following objects:
 
\begin{itemize}

\item ${\cal H}  $ : Hilbert space of the original problem.

\item ${\cal H }' $: Hilbert space of the effective degrees of
freedom.

\item $H $: Hamiltonian acting in $\cal H$.

\item $H'$: Hamiltonian acting in $\cal H'$ (effective Hamiltonian).

\item $T  $ : embedding operator : ${\cal H }'\longrightarrow {\cal
H}$

\item $T^{\dagger }  $ :truncation operator : ${\cal H}
\longrightarrow {\cal H}'$

\end{itemize}

The problem now is to relate $H$, $H'$ and $T$. The criterium to
accomplish this task is that  $H$ and $H'$ have in common their low
lying spectrum. An exact implementation of this  is given by the
following equation:

\begin{equation} H T = T
H'                                                   \label{3}
\end{equation}

\noindent which imply that if $\Psi '_{E'}$ is an eigenstate of $H'$
then  $T\Psi '_{E'}$ is an eigenstate of $H$ with the same eigenvalue
(unless it belongs to the kernel of $T$: $T\Psi '_{E'}=0$), indeed,

\begin{equation} HT\Psi '_{E'} =  T H'\Psi '_{E'} = E' T\Psi
'_{E'}                                                  \label{4}
\end{equation}

To avoid the possibility that $T\Psi '=0$ with $\Psi '\neq 0$, we
shall impose on $T$ the condition,

\begin{equation} T^{\dagger } T = 1_{\cal
H'}                                                   \label{5}
\end{equation}

\noindent such that 

\begin{equation} \Psi = T\Psi '  \Rightarrow \Psi ' = T^{\dag}
\Psi                                                  \label{6}
\end{equation}

\noindent Condition (\ref{5}) thus stablishes a one to one relation
between $\cal H'$ and Im($T$)  in $\cal H$.

\noindent Observe that Eq. (\ref{3}) is nothing but the commutativity
of the following  diagram:

\begin{center} \[ \begin{array}{llcll}
  & {\cal H'} & \stackrel{T}{\longrightarrow} & {\cal H}  &  \\
 H' & \downarrow &     &  \downarrow & H  \\
  &  {\cal H'}  &  \stackrel{T}{\longrightarrow} & {\cal H} & 
\end{array} \] \end{center}

 Eqs. (\ref{3}) and (\ref{5}) characterize what may be called exact
renormalization group method  (ERG) in the sense that  the whole
spectrum of $H'$ is mapped onto a part (usually the bottom part)  of
the spectrum of $H$. In practical cases though the exact solution of 
Eqs. (\ref{3}) and (\ref{5})  is not possible so that one has to
resort to approximations (see later on). Considering  Eqs. (\ref{3})
and (\ref{5}) we can set up the effective Hamiltonian $H'$ as:

 \begin{equation} H' = T^{\dag} H
T                                                   \label{10}
\end{equation}

\noindent This equation does not imply that the eigenvectors of  $H'$
are mapped onto eigenvectors  of $H$. Notice that Eq.(\ref{10})
together with (\ref{5}) does not imply Eq. (\ref{3}). This happens
because the converse of Eq.(\ref{5}), namely $TT^{\dag } \neq 1_{\cal
H}$  is not true, since otherwise this equation together with
(\ref{5}) would imply that the Hilbert spaces ${\cal H}$ and ${\cal H'}$
are isomorphic while on the other  hand the truncation inherent to
the RG method assumes that $dim {\cal H'} < dim {\cal H}.$

What Eq.(\ref{10}) really implies is that the mean energy of $H'$ for
the states $\Psi '$ of $\cal H'$  coincides with the mean energy of
$H$ for those states of $\cal H$ obtained through the embedding  $T$,
namely,

  \begin{equation} <\Psi '|H'|\Psi'> =  <T\Psi ' |H| T\Psi'>  
\label{11} \end{equation}

In other words $ T\Psi'$ is used as a variational state for the
eigenstates of the Hamiltonian $H$. In  particular $T$ should be
chosen in such a way  that the states truncated in $\cal H$ , which
go down  to $\cal H'$, are the ones expected to contribute the most
to the ground state of $H$. Thus Eq.  (\ref{10}) is the basis of the
so called variational renormalization group method (VRG)
 As a matter  of fact, the VRG method was
the first one to be proposed.  The ERG came afterwards as a
perturbative extension of the former (see later on).  

\noindent  More generally, any operator $\cal O$ acting in $\cal H$
can be ``pushed down" or renormalized to a new  operator $\cal O'$
which acts in $\cal H'$ defined by the formula,

  \begin{equation} {\cal O}'  = T^{\dag} {\cal O}
T                                       \label{43} 
\end{equation}

\noindent Notice that Eq.(\ref{10}) is a particular case of this
equation if choose {\cal O} to  be the Hamiltonian $H$. 

In so far we have not made use of the all important concept of the
block,  but a practical implementation of the VRG or ERG methods does
require it. The central role played by this concept  makes all the
real-space RG-methods to be block methods.

Once we have established the main features of the RG-program,  there
is quite a freedom to implement specifically these fundamentals. We may
classify this freedom in two aspects:

\begin{itemize}

\item The choice to how to reduce the size of the lattice.

\item The choice of how many states to be retained in the truncation
procedure.

\end{itemize}

\noindent We shall address the first aspect now. There are mainly two
procedures to reduce the  size of the lattice:

\begin{itemize}

\item by dividing the lattice into blocks with $n_B$ sites each. This
is the blocking method  introduced by Kadanoff to treat spin lattice
systems. See figure 1.

\item by retrieving site by site of the lattice at each step of the
RG-program. This is the procedure used by Wilson in his RG-treatment
of the Kondo problem. This method is clearly more suitable  when the
lattice is one-dimensional.

\end{itemize}

  We shall be dealing with the Kadanoff 
 block methods mainly because they are well suited to perform 
analytical computations and because they are conceptually easy to be
extended to higher  dimensions. On the contrary, the DMRG method
introduced by White \cite{white}  works with the Wilsonian numerical
RG-procedure what makes it intrinsically one-dimensional  and
difficult to be generalized to more dimensions. 
This situation has changed recently in part as S.R. White has devised
a numerical improvement of the DMRG which is applicable to a 1/5-depleted
2D lattice \cite{white2D}.
We have
formulated our  Variational and Fokker-Planck DMRG procedures as block
renormalization methods \cite{dm-germanyo}.

\noindent Block RG-methods have recently received also renewed 
attention in one-dimensional problems in connection to what is called
a {\em quantum group} symmetry \cite{q-germanyo}, \cite{qbis-germanyo}. 
Based upon this
symmetry we have constructed a new BRG-method that we call $q$-RG
which among other features it is able to predict the exact line of
critical XXZ models in the  Anisotropic Heisenberg model, unlike the
standard BRG-method.


\section{The Role of Boundary Conditions and Real-Space RG}

The first advance in trying to understand the sometimes bad numerical performance of the BRG
methods came in the understanding of the effect of {\em boundary conditions} (BC) on the 
standard RG procedure \cite{white-noack}.

White and Noack \cite{white-noack} pointed out that the standard BRG approach of neglecting all 
connections to the neighbouring blocks during the diagonalization of the block Hamiltonian $H_B$ 
introduces large errors which cannot be corrected by any reasonable increase in the number of 
states kept. Moreover, in order to isolate the origin of this problem they study an extremely simple 
model: a free particle in a 1D lattice. As a matter of fact, it was Wilson \cite{wilson-unp} 
who pointed out the importance of understanding real-space RG in the context of this simple 
tight-binding model where the standard BRG clearly fails as we are going to show.

\noindent The reason for this failure can be traced back to the importance of the boundary 
conditions in diagonalizing the states of a given block Hamiltonian $H_B$ in which the lattice is 
decomposed into.  Notice that in this fashion we are isolating  a given block from the rest of the 
lattice and this applies a {\em particular BC}  to the block. However, the block is not truly 
isolated! A statement which is the more relevant the more strongly correlated  is the system 
under consideration. Thus, if the rest of the lattice were there it would apply different  BC's to the 
boundaries of the block. This in turn makes the standard block-diagonalization conceptually not 
faithfully suited to account for the interaction with the rest of the lattice.

Once the origin of the problem is brought about the solution is also apparent: devise a method to 
change the boundary conditions in the block in order to mimick the interaction with the rest of 
the lattice. This is called the Combination of Boundary Conditions (CBC) method which yields 
very good numerical results. This method has not yet been generalized to interacting systems.
However in reference \cite{white} an alternative approach is proposed under the name of 
Density Matrix Renormalization Group (DMRG) which applies to more general situations and also 
produces quite accurate results. 

In a recent paper \cite{bc-germanyo}
 we have reconsidered again the role of BC's in the real space RG method for the 
case of a single-particle problem in a box. The continuum version of this Hamiltonian is 
simply $H = - \frac{\partial ^2}{\partial x^2}$. We shall consider open chains with two types of 
BC's at the ends:

\begin{equation}
\mbox{Fixed BC's:} \ \ \ \  \psi (0) = \psi (L) = 0           \label{bc1a}
\end{equation}

\begin{equation}
\mbox{Free BC's:}  \ \ \ \  \frac{\partial \psi}{\partial x} (0) =
 \frac{\partial \psi}{\partial x} (L) = 0           \label{bc1b}
\end{equation}

\noindent The lattice version of $H$ for each type of BC's is given as follows:

\begin{equation}
 H_{Fixed} = \left( \begin{array}{cccccc}
2 & -1 & & & & \\
 -1 & 2 & -1& &  & \\
 & -1 &2 & & &  \\
&  &  &  \ddots &  &  \\
 & &  & & 2 & -1  \\
 & &  & & -1 & 2       \label{bc2}
\end{array}              \right), \quad 
 H_{Free} = \left( \begin{array}{cccccc}
1 & -1 & & & & \\
 -1 & 2 & -1& &  & \\
 & -1 &2 & & &  \\
&  &  &  \ddots &  &  \\
 & &  & & 2 & -1  \\
 & &  & & -1 & 1      
\end{array}              \right)
\end{equation}

\noindent The only difference between $H_{Fixed}$ and $H_{Free}$ appear at the first and last 
diagonal entry ($2 \leftrightarrow 1$). The exact solution of (\ref{bc2})  is very well-known 
and we give it for completeness:

\begin{equation}
\mbox{Fixed BC's:} \ \ \ \  \psi_n (j) = N_n^{Fx} \sin \frac{\pi (n+1)}{N+1} j,\ \ 
E_n =  4 \sin ^2(\frac{\pi (n+1)}{2(N+1)})       \label{bc4a}
\end{equation}

\begin{equation}
\mbox{Free BC's:}  \ \ \ \  \psi_n (j) = N_n^{Fr} \cos \frac{\pi n}{N} (j - \frac{1}{2}), \ \ 
E_n =    4 \sin ^2(\frac{\pi n}{2N})      \label{bc4b}
\end{equation}
\[ j = 1,2,\ldots ,N; \ \ \ \ n = 0,1,\ldots ,N-1. \]

\noindent where the $N_n's$ are normalization constants
and $N$ is the number of sites of the chain.

Before getting into the problem of the renormalization of these Hamiltonians, 
it is worth to pointing 
out another physical realization of $H_{Free}$:
 {\em A simple magnon above a ferromagnetic
background satisfies Free BC's}. See \cite{bc-germanyo} for more details.

Now let us get to the problem of renormalizing the tight-binding Hamiltonians
 (\ref{bc2}).

In figure 2 we show the ground state and first excited 
states of the chain with fixed and 
free BC's.

It is clear from Fig.2 that {\em a standard Block RG method is not appropiate 
to study the ground state 
of fixed BC's since this state is non-homogeneous while the block truncation does not take into account this fact.}

\noindent Each piece of the ground state within each block satisfies BC's which vary from block to 
block. This is the motivation of reference \cite{white-noack} to consider different BC's  in the 
block method, yielding quite accurate results.

\noindent We observe that the standard RG method performs rather poorly as compared to the 
CBC method which yields quite the exact results.

\noindent The other alternative to the CBC method is the Density Matrix RG method which can 
be phrased by saying that the rest of the chain produces on every block the appropiate BC's 
to be applied to its ends, and it has the virtue that can be generalized to other models, 
something which is not the case as for the CBC method.

\noindent On the other hand, the ground state of $H_{Free}$ is an homogeneous state (see Fig.2) 
which in turn suggests that a standard RG analysis may work for this type of BC's. We shall 
show that this is indeed the case if the RG procedure is properly defined. 
The key of our 
RG-prescription is to notice that $H_{Free}$ has a geometrical meaning: {\em $H_{Free}$ is the
incidence matrix of a graph}, and it is called minus the discrete laplacian $-\Delta$ of that
graph. Notice that $H_{Fixed}$ has not such geometrical interpretation, in fact, it concides with 
the Dynkin diagram of the algebra $A_N$.

\noindent Based on this observation the Kadanoff blocking
 is nothing but the breaking of the graph 
into $N/n_s$ disconnected graphs of $n_s$ sites each. We shall choose $n_s = 3$ in our later 
computation.

The previous geometrical interpretation of $H_{Free}$ suggests that we choose the block 
Hamiltonian $H_B$ to be the incidence matrix  of a disconnected graph, namely,

\begin{equation}
 H_{B} = \left( \begin{array}{ccccccc}
1 & -1 & & & & & \\
 -1 & 2 & -1& &  &  & \\
 & -1 &1 & & &  & \\
&  &  &  1 & -1 &  & \\
 & &  &-1 & 2 & -1 &  \\
 & &  & & -1 & 1     &  \\
 & &  & &  &    & \ddots
 \label{bc5}
\end{array}              \right), \quad 
 H_{BB} = \left( \begin{array}{cccccccc}
0 &  & & & & & &\\
  & 0 & & &  &  & &\\
 &  &1 & -1& &  & &\\
&  &  -1&  1 &  &  & &\\
 & &  &  & 0 &  &  &\\
 & &  & &  & 1     &  -1 &\\
 & &  & &  & -1   & 1 & \\
 & &  & &  &    &  &\ddots
\end{array}              \right)
\end{equation}

\noindent andthe interblock Hamiltonian $H_{BB}$ above describes the interaction 
between blocks.

\noindent $H_{BB}$ in turn also coincides with 
the incidence matrix  of a graph which contains 
the missing  links  which connects consecutive blocks. In a few words:
our RG-prescription introduces free BC's at the ends of every block. This condition fixes uniquely
the breaking of $H_{Free}$ into the sum $H_B + H_{BB}$. This is the choice we make.
It should be emphasized that the
splitting of $H_{Free}$ into two parts $H_B + H_{BB}$
is by no means unique, so that different choices may lead to very different results.

\noindent Prior to any computation we notice that the previous RG-prescription should lead to an 
exact value of the ground state energy, for the ground state of each block is again a constant 
function. The question is therefore to what extent our method is capable of describing the 
excited states. We shall concentrate ourselves to the first excited state since computations 
can be carried out analytically.

First of all we diagonalize $H_B$ within each block of 3 sites,  keeping only the ground state
$\psi _0^{(0)}$ and the first excited state $\psi _1^{(0)}$ ($3 \rightarrow 2$ truncation). 
The superscript denotes the initial step in the truncation method.
In figure 3 we picture the 3 eigenvectors for the 3-site Hamiltonian which will be
the building blocks for our BRG, namely the two lowest ones.
In the standard RG method we would
choose $\psi _0^{(0)}$  and $\psi _1^{(0)}$ as the orthonormal basis for the truncated Hilbert 
space and obtain the effective Hamiltonian $H'_{B}$ and $H'_{BB}$. In our case it is convenient to 
express these effective Hamiltonians in a basis expanded by the following linear combination:

\begin{equation}
\psi _+^{(0)}  = \frac{1}{\sqrt{2}} (\psi _0^{(0)}  +  \psi _1^{(0)})                  \label{bc7a}
\end{equation}

\begin{equation}
\psi _-^{(0)}  = \frac{1}{\sqrt{2}} (\psi _0^{(0)}  -  \psi _1^{(0)})                  \label{bc7b}
\end{equation}

\noindent which are also an orthonormal basis of the truncated Hilbert space.

\noindent In this basis the truncation of $H_B$ reads as follows,

\begin{equation}
 H_B \longrightarrow H'_{B} = \left( \begin{array}{ccccc}
A & 0 & 0 & & \\
  0 & A & 0 & &  \\
 0 &  0 & A & & \\
&  &  &  \ddots &  \\
 & &  &  & A\\
 \label{bc8}
\end{array}              \right), \quad 
A = \frac{\epsilon}{2}   \left( \begin{array}{cc}
1 & -1  \\
  -1 & 1   \\
\end{array}              \right)
\end{equation}

\noindent with $\epsilon$ taking on the value $\epsilon^{(0)} = 1$ in the initial step of the RG-method,
which is the energy of the 
state $\psi _1^{(0)}$.

The truncation of $H_{BB}$ is more complicated, the result being:

\begin{equation}
 H_{BB} \longrightarrow H'_{BB} = \left( \begin{array}{cccccc}
B & C & 0 & & & \\
  C^t & D + B & C & &  & \\
 0 &  C^t & D + B & & &  \\
&  &  &  \ddots &   & \\
&  &  &  D + B & C  \\
 & &  &  C^t & D \\
 \label{bc9}
\end{array}              \right)
\end{equation}

\[ 
B =   \left( \begin{array}{cc}
a^2 & a b  \\
  a b & b^2   \\
\end{array}              \right) \ \ 
C =   \left( \begin{array}{cc}
-a b & -a^2  \\
  -b^2 & -a b   \\
\end{array}              \right) \ \ 
D =   \left( \begin{array}{cc}
b^2 & a b  \\
  a b & a^2   \\
\end{array}              \right)
\]

\noindent  with $a$ $b$ taking on the values $a^{(0)} = \frac{1}{\sqrt{6}} - \frac{1}{2}$ and 
$b^{(0)} = \frac{1}{\sqrt{6}} + \frac{1}{2}$ in the initial step of the RG-method.

The nice feature about the basis (\ref{bc7a})-(\ref{bc7b})
 is that all rows and columns of (\ref{bc8}) and (\ref{bc9}) 
add up to zero, just like the original Hamiltonians (\ref{bc5}), implying that the 
constant vector is an eigenvector with zero eigenvalue of the renormalized Hamiltonian!

We shall call $H_{N/3}(\epsilon, a, b)$ the sum of the Hamiltonians 
(\ref{bc8}) and (\ref{bc9}) for 
generic values of $\epsilon$, $a$ and $b$. Next step in our RG-procedure is to form blocks 
of 4 states of the new Hamiltonian $H_{N/3}(\epsilon, a, b)$ and truncating to the two lowest 
$\psi _0^{(1)}$ and  $\psi _1^{(1)}$ energy states within each 4-block (4 $\rightarrow$ 2 truncation). 
The reason for this change in the number of sites per block (from 3 to 4) 
 is motivated by the form of $H'_{BB}$ in (\ref{bc9}) and the fact that if we try to make a second step
in the RG-method with 3-blocks the method is doomed 
to failure because the constant 
state of Fig.2 would no longer be the ground state.

\noindent Fortunately enough, with 4-blocks if we define new states 
$\psi _+^{(1)}$ and  $\psi _-^{(1)}$ in the same form as we did in Eq.(\ref{bc2}), we obtain that 
the new effective Hamiltonian is obtained by a redefinition of the parameters, namely,

\begin{equation}
H_{N/3}(\epsilon, a, b)  \longrightarrow H_{N/6}(\epsilon', a', b') \label{bc10a}
\end{equation}

\begin{equation}
\epsilon' = \frac{\epsilon}{2} + a^2 +b^2 - \Delta  \label{bc10b}
\end{equation}

\begin{equation}
a' = \frac{1}{2 \sqrt{2}} \left[ a + b - 
\frac{a (a^2 - 3 b^2 + \Delta) + \frac{b \epsilon}{2}}
{\sqrt{\Delta (\Delta + a^2 - b^2)}}   \right]  \label{bc10c}
\end{equation}

\begin{equation}
b' = \frac{1}{2 \sqrt{2}} \left[ a + b + 
\frac{a (a^2 - 3 b^2 + \Delta) + \frac{b \epsilon}{2}}
{\sqrt{\Delta (\Delta + a^2 - b^2)}}   \right]  \label{bc10d}
\end{equation}

\begin{equation}
\Delta \equiv \sqrt{(a^2 - b^2)^2 + (\frac{\epsilon}{2}  - 2 a b)^2} \label{bc10e}
\end{equation}

\noindent In this fashion, the constant state of Fig.2 is again the ground state of the model and 
moreover, upon iteration of Eqs.(\ref{bc10a})-(\ref{bc10e}) there are no level crossing among the excited 
states.
Otherwise stated this means that the level structure of the block Hamiltonian $H_B$ is preserved 
under the action of our BRG-method based upon the reduction from 4 to 2 states.

The energy $E_1(N)$ of the first excited state of a chain with $N = 3 \times 2^m$ sites can be 
obtained iterating $m$ times Eqs.(\ref{bc10a})-(\ref{bc10e}):

\begin{equation}
E_1(N=3 \times 2^m) \equiv \epsilon^{(m)}  \label{bc11}
\end{equation}

\noindent The initial data are given by:

\begin{equation}
\epsilon^{(0)} = 1, \ \
a^{(0)} = \frac{1}{\sqrt{6}} - \frac{1}{2},\ \  
b^{(0)} = \frac{1}{\sqrt{6}} + \frac{1}{2}            \label{bc12}
\end{equation}

\noindent For low values 
of $N$ the deviation of $\epsilon ^{(m)}$ with respect to the exact result is small
(see \cite{bc-germanyo}).
Recall that we are only keeping two states in our RG-procedure, and that the ground state 
energy is exactly zero by construction!. But what is more interesting about these 
(see \cite{bc-germanyo}) is that we are able to obtain the correct size dependence, i.e., $1/N^2$ of $\epsilon ^{(m)}$.
As a matter of fact, the energy of the first excited state behaves for large $N$ as (\ref{bc4b}):

\begin{equation}
E_1^{(exact)}(N) \sim c_{exact}/N^2, \ \ \ \ \mbox{with} \ \ c_{exact} = \pi^2          \label{bc13}
\end{equation}

\noindent while our BRG-method gives,

\begin{equation}
E_1^{(BRG)}(N) \sim c_{BRG}/N^2, \ \ \ \ \mbox{with} \ \ c_{BRG} = 12.6751         \label{bc14}
\end{equation}

The achievement of the $1/N^2$-law is a remarkable result which in turn allows us to match the 
correct order of magnitude of the energy. For instance, for 10 iterations our RG-method with 2 states kept 
gives  the energy of the 
order of $10^{-6}$, which is precisely the same order of magnitude as for the CBC method 
but with 8 states kept in the case of Fixed BC's. Recall that the standard BRG 
performs as bad as a $10^{-2}$ order of magnitude.

\section{Wave-Function Reconstruction}

Insofar we have only used our RG method to ``reconstruct" the energies of the
lowest states step by step, but we can also use this method to reconstruct the
shape of the associate wave function in the real space. This simple fact leads to
the consideration of RG applications beyond the original scope for which it was 
devised. As in this fashion we are making a picture of the wave function, it is natural
to use the RG as a image compression method for coding images in order to 
facilitate its transport through the networks. For more details see the notes by
Nishino in these proceedings \cite{nishinonote}.

We are able to make a reasonable picture of the first excited state 
wave-function based upon our BRG-procedure when compared with the exact form 
depicted in Fig.2. As we are working with a real-space realization of the renormalization 
group method, this is something we have at hand. To do this we need to perform a ``reconstruction" 
of the wave-function. This reconstruction amounts to plot the form of our aproximate 
wave-function in each and every  of the 3-blocks out of the $2^{m+1}$ in which the 
original chain is decomposed into under the BRG-procedure. Recall that in the initial step we 
started out with blocks of 3 states keeping the two lowest states $\psi_0^{(0)}$ and $\psi_1^{(0)}$ 
($3 \rightarrow 2$ truncation). 
In the next step we make blocks of 4 states keeping the two lowest 
states $\psi_0^{(1)}$ and $\psi_1^{(1)}$ 
($4 \rightarrow 2$ truncation) and then we perfom the iteration procedure over and over. As a 
result of this procedure we may express the two lowest wave functions of the $m+1$-th step 
in terms of those of the previous $m$-th step by means of the following matricial form:

\begin{equation}
\left( \begin{array}{c}
\psi_0^{(m+1)}  \\
  \psi_1^{(m+1)} \\
\end{array}              \right)  =   \frac{1}{\sqrt{2}}  \left( \begin{array}{cc}
1 & 0  \\
  \alpha_m & \beta_m  \\
\end{array}              \right)  \left( \begin{array}{c}
\psi_0^{(m)}  \\
  \psi_1^{(m)} \\
\end{array}              \right)_L  + 
 \frac{1}{\sqrt{2}}  \left( \begin{array}{cc}
1 & 0  \\
  -\alpha_m & \beta_m  \\
\end{array}              \right)  \left( \begin{array}{c}
\psi_0^{(m)}  \\
  \psi_1^{(m)} \\
\end{array}              \right)_R  \label{bc15}
\end{equation}

\noindent where the LHS of Eq. (\ref{bc15}) represents the wave function of $3\times 2^{m+1}$ sites 
while in the RHS we have a left-wave-function of $3\times 2^{m}$ sites and another 
right-wave-function of $3\times 2^{m}$ sites, so that everything squares. The parameters 
appearing in Eq. (\ref{bc15}) turn out to be given by:

\begin{equation}
 \alpha_m = \frac{(\frac{\epsilon_m}{2} - 2 a_m b_m) + (a_m^2 - b_m^2 + \Delta_m)}
{2 \sqrt{\Delta_m (\Delta_m + a_m^2 - b_m^2)}}     \label{bc16}
\end{equation}

\begin{equation}
 \beta_m = \frac{(\frac{\epsilon_m}{2} - 2 a_m b_m) - (a_m^2 - b_m^2 + \Delta_m)}
{2 \sqrt{\Delta_m (\Delta_m + a_m^2 - b_m^2)}}     \label{bc17}
\end{equation}

\noindent with $\Delta_m$ as in Eq.(\ref{bc10e}). Their initial values are $\alpha_0=1/\sqrt{10}$
and $\beta_0=3/\sqrt{10}$.

\noindent We may recast Eq. (\ref{bc15})  in more compact form by writing:

\begin{equation}
 \Psi^{(m+1)} = L_m  \Psi^{(m)}_L  +   R_m  \Psi^{(m)}_R   \label{bc18}
\end{equation}

\noindent where 

\begin{equation}
 L_m  =  \frac{1}{\sqrt{2}}  \left( \begin{array}{cc}
1 & 0  \\
  \alpha_m & \beta_m  \\
\end{array}              \right), \quad     
 R_m  =  \frac{1}{\sqrt{2}}  \left( \begin{array}{cc}
1 & 0  \\
  -\alpha_m & \beta_m  \\
\end{array}              \right)    
\label{bc19}
\end{equation}

\noindent We may call Eq.(\ref{bc18}) the {\em reconstruction equation}. This is the master equation 
that when iterated ``downwards" (reconstruction) allows us to obtain the picture of our 
approximate BRG-wave-function corresponding to every and each block of 3 sites of the 
$2^{m+1}$ blocks in which the chain is decomposed into. At the end of the iteration procedure
we end up with expressions for the values of the 3-sites wave-functions in terms of the 
initial two lowest states $\psi_0^{(0)}$ and $\psi_1^{(0)}$. The first one is a constant function 
while the second is a straight line of negative slope. Thus, these two states turn out to be the 
building blocks of our BRG-procedure.

When using the reconstruction equation to obtain the wave function we may use a binary code 
based upon the labels L (left) and R (right) to keep track of the different 3-sites blocks which 
make up the chain. Thus, in one dimension the RG-blocks are in a one-to-one correspondence  
with a binary numerical system. In general, for other dimensions we may state 
squematically the following 
correspondence:

\[ \mbox{BRG-prescription} \longleftrightarrow \mbox{``Number System"} \]

\noindent This simple observation is the basis of a coding system for compressing
pictures whatever may be its origin. In two dimensions we need more digits to
make the coding, but it works likewise and serves for image compression 
\cite{nishinonote}.


\section{The Correlated Block Renormalization Group (CBRG)}

 We have already mentioned in the Introduction
 that the DMRG is not only a powerful 
computational method but also a source of inspiration for further 
works concerning the RG.
For these reasons, it is worthwhile to explore different options
or alternatives to the DMRG which may be useful in situations where 
the DMRG encounters difficulties, as in the case of 2D quantum systems.
The main message of the DMRG is that blocks are correlated. The 
implementation of this idea by means of the 
density matrix formalism may 
be not the unique way to proceed. On the other hand, the 
``onion-scheme" a la Wilson adopted by the DMRG, while being one of 
the reasons of its spectacular accuracy, imposes certain limitations.

\noindent At this stage it is not clear how fundamental are the 
density-matrix formalism or the onion-scheme for a RG method which
takes into account the correlation between blocks.
One can indeed combine the Kadanoff block method with the use of a
density matrix in the process of truncation, as in reference 
\cite{dm-germanyo}. More work remains to be done to see wheather there
is a real improvement of the standard BRG method by combining it with
the DMRG as in \cite{dm-germanyo}.
In this section we want to explore another possibility which is to give
up both the density matrix and the onion-scheme (see \cite{cbrg}).
With this point of view in mind, it would seem that we should come 
pretty close to the standard BRG method, were it not for the enormous 
freedom hidden in a Real-Space RG method. This freedom comes from 
the separation of the Hamiltonian into an intrablock $H_B$ and an 
interblock $H_{BB}$ Hamiltonian. 
This is a source of ambiguities which can be 
sometimes mitigated with the aide of symmetry arguments, but not 
fully eliminated though. This ambiguity shows up specially for
terms in the Hamiltonian acting at the boundaries of the block.
There are no general criteria as to how to include this type of terms
either into the intrablock or into the interblock Hamiltonians, or 
into both! For example, in the 1D Ising model in a transverse field
(ITF model), a choice which preserves the selfduality of the model
attributes some self-couplings to the $H_B$ and others to the 
interblock $H_{BB}$, and it yields to an exact value of the critical 
point and the critical exponent $\nu$ \cite{fpacheco}, 
\cite{q-germanyo}. The ambiguity in the splitting of $H$ into the 
sum $H_B + H_{BB}$ thus affect deeply the truncation procedure itself,
which is based on the diagonalization of $H_B$. Rather than blaming
the BRG for its lack of uniqueness, we should use its freedom to 
allow the blocks to become correlated in the RG procedure. In our
present approach this correlation will be taken into account in a 
``dynamical" way rather than in a ``statistical" way as in the 
DMRG. This will be achieved by the introduction of interblock operators
which reflect the ``influence" between neighbour blocks and which
are defined at the boundary of the block in the first step of 
our CBRG method.

We have chosen to illustrate our approach the 1D and 2D tight-binding
models mainly for simplicity reasons, but we believe that our 
method could be applied to more complicated problems. In fact, the 
first step in this direction was already undertaken in reference
\cite{bc-germanyo}, where only 2 states at each stage of the 
RG-blocking were retained. This in turn allowed us to obtain the 
$1/N^2$ scaling law for the size dependence of the 
first-excited-state energy.

 We shall give the general mathematical structure
underlying the results of reference \cite{bc-germanyo}.
This will allow us to retain more than two states in the RG-truncation
and also to consider the two-dimensional tight-binding model.
In this fashion, we shall recover the $n^2/N^2$ scaling law for the
$n$-th excited state of the 1D model and the scaling law 
$\frac{n^2_1+n^2_2}{N^2}$ in the 2D case. These results will then 
show that the CBRG method describes correctly the low energy behaviour
of the 1D and 2D Laplacian.

\section{The CBRG Method: One Dimension}

The problem  we want to study is the one-dimensional Tight-Binding
model in an open chain with different boundary conditions at its ends.
The Hamiltonian for this system takes the following matricial form,

\begin{equation}
 H_{b,b'} = \left( \begin{array}{cccccc} b& -1 & & & & \\
 -1 & 2 & -1& &  & \\
 & -1 &2 & & &  \\ &  &  &  \ddots &  &  \\
 & &  & & 2 & -1  \\
 & &  & & -1 & b'       \label{cb1} \end{array}              \right)
\end{equation}

\noindent where $b$ and $b'$ take on the values $1$ (or $2$)
corresponding to  Free (or Fixed) BC's respectively. This Hamiltonian
is the discrete version of the Laplacian $H=-\partial^2_x$, while the
Free or Fixed BC's correspond in  the continuum to the vanishing of
the wave function (Fixed BC's) or its spatial  derivative (Free BC's)
at the ends of the chain, i.e.,

\begin{equation}
 \begin{array}{cccc}  b = 2& \Rightarrow & \Psi (0) = 0 & \mbox{Fixed
BC} \\ b = 1& \Rightarrow & \frac{\partial\Psi}{\partial x}(0) = 0
& \mbox{Free BC} 
  \label{cb2} \end{array}              \end{equation}

\noindent and similarly for $b'$ which contains the BC at the other
end of the  chain.

\noindent Hence, altogether there are 4 Hamiltonians of the type in
(\ref{cb1}), whose  eigenstates and eigenvalues are the subject of our RG-techniques.

The first step in the RG method is to divide the lattice into blocks
containing  $n_s$ sites each and labeled with and index $p$
($=1,\ldots,N/n_s$). Let us suppose for a moment that we isolate the
$p$th-block from the rest of the lattice so that its dynamics, as an
independent entity, is governed by a Hamiltonian denoted by $A_p$,
which we may call {\em uncorrelated block Hamiltonian}. The
restoration of the block back into the lattice involves two effects.
The first one is that the BC's of the $p$-th block may change  under
the influence of the $p+1$ and $p-1$ blocks. We describe this change
of BC's by the action of  {\em Boundary Operators} denoted by
$B_{p,p\pm 1}$ on the $p$th-block. The second effect is  the 
interaction between the $p$th-block and its neighbours $p+1$ and
$p-1$,
 given by interaction  Hamiltonians $C_{p,p\pm1}$ which act
on both $p$ and $p+1$ blocks simultaneously. If the problem  under
consideration is translationally invariant, all the Hamiltonians
defined above are independent of the block label $p$, in which case
we denote them by,

\begin{equation}
 \begin{array}{cc}  A_p = A & \\ B_{p,p+1} = B_R& B_{p,p-1} = B_L \\
C_{p,p+1} = C & C_{p,p-1} = C^{\dagger}
  \label{cb3} \end{array}              \end{equation}

The $H_{Free, Free}$ Hamiltonian (\ref{cb1}) gives an example of
this as we shall show below. Hence, for the time being, we shall
consider the situation described by (\ref{cb3}) and leave the more
general case after explaining the general ideas.

In the standard BRG method the block Hamiltonian $H_B$ and the
interblock  Hamiltonian $H_{BB}$ are given,
according to our  previous definitions,
 by the following formulas

\begin{equation} H_B = A + B_L + B_R  \label{cb4a}            
\end{equation}

\begin{equation}
 H_{BB} =  \left( \begin{array}{cc}  0& C\\ C^{\dagger} &0
  \label{cb4b} \end{array}      \right)         \end{equation}

\noindent The whole Hamiltonian is by all means the sum of $H_B$ and
$H_{BB}$ for all the blocks of the chain.

\noindent Next step in the RG method is to diagonalize $H_B$ and keep
its, say $m$ ($m<n_s$), lowest eigenstates. The truncation is given by a
$n_s \times m$ matrix $T$ whose columns are precisely the components
of the $m$ lowest  eigenstates of $H_B$. The renormalized Hamiltonian
in the new basis is given by,

\begin{equation} H' = T^{\dagger} (H_B + H_{BB}) T  \label{cb4c}            
\end{equation}

\noindent At first sight from Eq. (\ref{cb4a}) it would seem that we
have taken into account the effect of the BC's on a given block.
However, as the examples show, this is quite a bit illusory. On the
other hand, the distinction among $A$, $B_L$  and $B_R$ is rather
inmaterial as far as $H_B$ is concerned, and in fact  no distinction
of this sort is made in the standard BRG formalism. Finally, let us
observe that $H_B$ and $H_{BB}$ play rather different roles in the
truncation procedure. This asymmetry has been observed as a source of
problems by several authors in the past \cite{fpacheco}, \cite{rabin}.

We shall mention that
 this asymmetry has recently been related to quantum groups in a
fashion  which has led to a new RG method called the Renormalization
Quantum  Group method \cite{q-germanyo}, \cite{qbis-germanyo}.

Therefore, from various points of view, one is urged to make more
explicit the role played by the BC-operators $B_L$ and $B_R$ in our
CBRG procedure. For this purpose, we have found convenient to use the
concept of superblock already introduced in reference
\cite{white-noack}. We shall define a superblock as the set of two
consecutive blocks, $p$ and  $p+1$ and denoted by $(p,p+1)$. The great
advantage of the superblock is that it allows us to  materialize the 
distinction among $A$, $B_L$  and $B_R$ . In fact, just as the
isolation of a single block leads us to the definition of the
Hamiltonian $A$, the isolation  of two blocks contained in a
superblock  allows us to define  $B_L$, $B_R$ and also $C$ through
the superblock  Hamiltonian $ H_{sB} $ as follows,

\begin{equation}
 H_{sB} =  \left( \begin{array}{cc}  A + B_R& C\\ C^{\dagger}
 & A + B_L
  \label{cb5} \end{array}      \right)         \end{equation}

\noindent Similarly, the Hamiltonian describing the interaction
between superblocks is given by (see Fig.4)

\begin{equation}
 H_{sB,sB} =  \left( \begin{array}{cccc}  
0& & & \\ 
& B_R& C&\\ 
&C^{\dagger} &  B_L& \\ 
& & &0
  \label{cb6} \end{array}      \right)         \end{equation}

\noindent Now instead of diagonalizing $H_B$ in Eq. (\ref{cb4a}), in the
CBRG method we shall diagonalize $H_{sB}$ in Eq. (\ref{cb5}), and
afterwards keep
 the $m=n_s$  lowest eigenstates in the tight-binding  model.
As in the standard BRG method,  the change to the truncated basis
defines the renormalized operators as follows:

\begin{equation} H_{sB} \longrightarrow T^{\dagger} H_{sB} T = A'
\label{cb7a}             \end{equation}

\begin{equation} H_{sB,sB} \longrightarrow T^{\dagger}
H_{sB,sB} T=  \left(
\begin{array}{cc}  B'_R& C'\\ \mbox{$C'$}^{\dagger} &B'_L
  \label{cb7b} \end{array}      \right)         \end{equation}

\noindent where the matrices $A'$,$B'_R$, $B'_L$ and $C'$ are the
renormalized version of the operators $A$,$B_R$, $B_L$ and $C$, and
they exhibit the same  geometrical interpretation for the
renormalized block as their unprimed partners for the original
blocks. 

If we set $B_R = B_L = 0$ in Eqs. (\ref{cb5}) and (\ref{cb6}), then after
the first RG-step we get $B'_R = B'_L = 0$ and thus the previous
RG-scheme coincides with the  standard BRG. We may say that
uncorrelated blocks are in a sense  a fixed point of our method.
However, this fixed point may be unstable, and to explore this 
possibility
one has to look for non-vanishing $B$-operators and their 
RG-evolution.

Let us address now some examples. We shall first study the
Hamiltonian (\ref{cb1}) with Free BC's at the ends ($b=b'=1$).
Choosing $n_s=3$ for example, we see that the choice for the 
operators $A$,$B_R$, $B_L$ and $C$ in the first step of the CBRG
procedure is given by,

\begin{equation} A =  \left( \begin{array}{ccc}  
 1& -1 &0 \\ -1 & 2
& -1\\
 0 & -1 & 1
   \end{array}      \right)  , \  B_R =  \left( \begin{array}{ccc}  
0&  & \\
 & 0 & \\
  &  & 1
   \end{array}      \right)  , \  B_L =  \left( \begin{array}{ccc}  
1&  & \\
 & 0 & \\
  &  & 0
   \end{array}      \right)  , \  C =  \left( \begin{array}{ccc}  
0& 0 &0 \\ 0 & 0 & 0\\
 -1 & 0 & 0
   \end{array}      \right)   \label{cb8}
      \end{equation}

\noindent This choice is equivalent to the assumption that an
isolated block satisfies Free BC's at its ends. The role of $B_R$
and $B_L$ is to {\em join} these blocks into a single chain. This is
the  {\em geometrical} explanation of Eqs. (\ref{cb8}). In more
general cases one must have to explore which is the best  choice.
The generalization of Eqs.(\ref{cb8}) to blocks with more  than 3
sites is obvious. In Table 1 we collect our CBRG-results for the first 5
excited states for a chain of $N=12 \times 2^6=768$ sites. 
Comparison with exact results gives a good agreement.

 An important feature of our CBRG method is that the 
$n^2/N^2$- scaling 
law ($N
\longrightarrow \infty$) for the energy of the $n$-excited states of
a chain made up of $N$ sites, is reproduced correctly (see Fig.5). 
In Table 2 we show the variation of the first-excited-state energy
with the size $N$ of the chain. From those values we can extract the 
corresponding $1/N^2$-law which turns out to be,

\begin{equation}
 E^{(CBRG)}_1 (N) = c^{(1)}_{CBRG} \frac{1}{N^2}, \ \ c^{(1)}_{CBRG} = 9.8080, 
 \ (N \longrightarrow \infty) \ \mbox{Free-Free BC's}\label{cb8a} 
  \end{equation}

\noindent while the exact value for the proportionality constant $c$ 
is $c_{\mbox{exact}}=\pi^2=9.86$. This amounts to a 0.6 \% error.

\noindent Likewise, we have  enough data so as to obtain the
corresponding $n^2/N^2$-law for the whole set of 5 excited states.
Thus, the scaling law we obtain is,

\begin{equation}
 E^{(CBRG)}_n (N) = c_{CBRG} \frac{n^2}{N^2}, \ \ c_{CBRG} = 8.4733, 
 \ (N \longrightarrow \infty) \ \mbox{Free-Free BC's}\label{cb8ab} 
  \end{equation}

\noindent which now amounts to a 7.34 \% error.
This is a natural fact from the worse knowledge of the highest excited states
 of the spectrum in a RG-scheme.

We can make even more explicit the successful achievement of the 
$1/N^2$-scaling law by leaving as a free adjustable parameter the
exponent of $1/N$ in addition to the proportionality constant.
Let us denote by $\theta $ this critical exponent. Using data from 
20 to 50 steps of our CBRG-method for several truncation of states
according to our scheme $2 n_s \rightarrow n_s$ 
(namely, $n_s=10,13,20$)
we arrive at the following results,

\begin{equation}
 E^{(CBRG)}_1 (N) = c_{CBRG} \frac{1}{N^{\theta}}, \ \ 
 \ (N \longrightarrow \infty) \ \mbox{Free-Free BC's}\label{cb8abc} 
  \end{equation}

\begin{equation} 
 \begin{array}{ccc}   
\mbox{For}\ \  20 \longrightarrow 10&  &\theta = 1.9708 \\
 \mbox{For}\ \  26 \longrightarrow 13&  &\theta = 1.9734 \\
 \mbox{For}\ \ 40 \longrightarrow 20&  &\theta = 1.9854 
   \end{array}      
 \label{cb8d}
      \end{equation}

\noindent These results clearly support the fact that we have
correctly reproduced the exact value of $\theta = 2$ for the 
finite-size critical exponent.

\noindent Last, but not least, as was proved in \cite{bc-germanyo} our
CBRG method gives the {\em exact} energy of the ground state for every
step of the RG-procedure for Free-Free BC's.

\noindent In tables 1 and 2 we also show the results we have obtained with a DMRG
analysis following White's method \cite{white}. This analysis is based on the 
onion scheme of enlarging the lattice site by site \'a la Wilson. The results coincide 
with the exact values within the 4 digits precision used here, but they start differing
when keeping more digits. Nevertheless, the DMRG is much more time consuming
than our CBRG method for it has to build the lattice site by site, while the CBRG 
reproduces the lattice by blocking which is much more efficient as far as CPU time
is concern, and moreover, it applies to two-dimensional situations where the onion
scheme  fails to reproduce the lattice. We have also performed the DMRG analysis
in 1D for Fixed-Fixed BC's in table 4 where the same considerations apply.The 
CBRG method is also more suitable for analytic formulations \cite{bc-germanyo}.

\noindent In reference \cite{bc-germanyo} it was shown that one can 
reproduce easily the wave function of the excited states. This 
procedure was called {\em reconstruction} since it works ``downwards"
in the CBRG method. The basic equation to be used is the  {\em
reconstruction equation} \cite{bc-germanyo},

\begin{equation} \Psi^{(r+1)} = L_r \Psi^{(r)}_L + R_r
\Psi^{(r)}_R\label{cb9} 
  \end{equation}

\noindent where $\Psi^{(r)}$ denotes the collection of $m$ lowest
eigenstates in the $r$-step of the CBRG-procedure, and $L_r$, $R_r$
are the block matrices in terms of which the truncation matrix 
$T^{\dagger}$
can be written as $T^{\dagger} = (L_r,R_r)$.

\noindent Our results for a chain of $N=12 \times 2^6=768$
sites and $n_s=6$ states kept are given in 
Fig.6 where we have plotted the first 5 excited states and compare them
with the exact wave functions. There are some remarkable facts regarding
these figures.
Firstly,  the number of nodes is correctly preserved by our CBRG wave 
functions.
Secondly, the Free-Free type of boundary conditions are also correctly reproduced
at the ends of the chain.
And lastly, it is worthwhile to point out that the CBRG wave functions
``degrade gracefully" as the energy of the excited state raises in
accordance with the fact that the lower the energy is, the more 
reliable are the results.

This ends the results for the Free-Free BC's. In order to address
other types of BC's we must come back to the case where the 
matrices $A$,$B_R$, $B_L$ and $C$ depend on each particular block.
Thus, for example, for the Fixed-Free BC's we shall choose as  the
uncorrelated $A$-matrix for the block located to the left end  of
the chain the following form ($n_s=3$),

\begin{equation} A_1 =  \left( \begin{array}{ccc}   2& -1 &0 \\ -1 &
2 & -1\\
 0 & -1 & 1
   \end{array}      \right) \ \ \mbox{Fixed-Free BC's} \label{cb10}
      \end{equation}

\noindent while the remaining matrices $A_p$, $(p=2,\ldots,N/3)$,
will be given by Eqs.(\ref{cb3}), (\ref{cb8}).

For Free-Fixed BC's, it is the last $A$-matrix which we have to take
different from the others, namely,

\begin{equation} A_{N/3} =  \left( \begin{array}{ccc}   1& -1 &0 \\
-1 & 2 & -1\\
 0 & -1 & 2
   \end{array}      \right) \ \ \mbox{Free-Fixed BC's} \label{cb11}
      \end{equation}

\noindent As for the Fixed-Fixed BC's case, we must change the $A$-matrix
at both ends of the chain according to the following prescription,

\begin{equation} A_1 =  \left( \begin{array}{ccc}   2& -1 &0 \\ -1 &
2 & -1\\
 0 & -1 & 1
   \end{array}      \right), \ \ A_{N/3} =  \left( \begin{array}{ccc}   1& -1 &0 \\
-1 & 2 & -1\\
 0 & -1 & 2
   \end{array}      \right) \ \ 
 \mbox{Fixed-Fixed BC's} \label{cb11a}
      \end{equation}

\noindent Then we follow the same steps as for the Free-Free BC's,
taking care that the  $A$,$B_R$, $B_L$ and $C$ matrices in each
CBRG-step may depend on  the position of the blocks. This implies
in particular that the embedding $T$-matrices may also vary from
block to block.

 In Tables 3 and 4 we  summarize our results for the Free-Fixed
and Fixed-Fixed BC's (Fixed-Free BC's are equivalent to Free-Fixed
BC's by parity transformation).
In these tables we present our CBRG results for the first 6 lowest lying states
for the 1D tight-binding model in a chain of $N=12 \times 2^5=384$ sites
with mixed boundary conditions, and they are compared against the exact
and standard BRG values.
Several remarks are in order.
First, we observe that the CBRG method produces  a good agreement with the 
exact results and certainly much more accurate by several orders of 
magnitude than the old BRG method.
Second, the CBRG method is able to reproduce the corresponding 
$n^2/N^2$-scaling 
laws for the spectrum of excited states in each case of mixed BC's.
Namely,

\begin{itemize} 

\item For Free-Fixed BC's and considering just the ground state, we have

\begin{equation}
 E^{(CBRG)}_0 (N) = c^{(0)}_{CBRG} \frac{1}{4N^2}, \ \ 
c^{(0)}_{CBRG} = 9.072, 
 \ (N \longrightarrow \infty) \ \mbox{Free-Fixed BC's}\label{cb11b} 
  \end{equation}

\noindent which amounts to a 8 \% error with respect to the exact value of 
$c_{exact}=\pi^2$.

\noindent As for the corresponding law for the whole spectrum, we find

\begin{equation}
 E^{(CBRG)}_n (N) = c_{CBRG} \frac{(n+1)^2}{4N^2}, \ \ 
c_{CBRG} = 7.6729, 
 \ (N \longrightarrow \infty) \ \mbox{Free-Fixed BC's}\label{cb11c} 
  \end{equation}

\noindent which represents a 11.5 \% error with respect to the exact value of
$\pi^2$.

\item For Free-Fixed BC's and considering just the ground state,
 we have

\begin{equation}
 E^{(CBRG)}_0 (N) = c^{(0)}_{CBRG} \frac{1}{N^2}, \ \ 
c^{(0)}_{CBRG} = 8.35, 
 \ (N \longrightarrow \infty) \ \mbox{Fixed-Fixed BC's}\label{cb11ddd} 
  \end{equation}

\noindent which amounts to a 8 \% error with respect to the exact value of 
$c_{exact}=\pi^2$.

\noindent As for the corresponding law for the whole spectrum, we find

\begin{equation}
 E^{(CBRG)}_n (N) = c_{CBRG} \frac{(n+1)^2}{N^2}, \ \ 
c_{CBRG} = 6.9696, 
 \ (N \longrightarrow \infty) \ \mbox{Fixed-Fixed BC's}\label{cb11dd} 
  \end{equation}

\noindent which represents a 16 \% error with respect to the exact value of
$\pi^2$.

\end{itemize}

\noindent We obtain bigger errors in the determination of these
scaling laws as compared with the Free-Free case mainly because
we have used less data in our fitting. Nevertherless, we find a 
good agreement with the exact results. Yet, there is another reason
as to why the accuracy in the case of mixed BC's is worse, namely,
the ground state wave function $\Psi_0$ is not homogeneous in space
as it is in the Free-Free case \cite{bc-germanyo}.
This makes the RG-procedure more involved and a source of extra 
uncertainties.

\noindent Let us mention in passing that we are also able to 
make a wave function
reconstruction in the mixed BC's cases as has been done for 
the Free-Free BC
case.

The outcome of all the results presented so far is that we 
have succeded in 
devising a Real-Space RG method capable of reproducing the 
correct eigenvalues
and eigenstates for the tight-binding model as originally 
envisaged by Wilson, within a certain accuracy which can in 
principle be improved.

Althoug the model we have employed to test our CBRG-method is a 
tight-binding model, there are some remarkable facts regarding
the fixed-point structure of our CBRG-solution that we would like
to stress.
Namely, we have found that after enough number of CBRG-iterations,
the matrices $A$, $B_L$, $B_R$ and $C$ in the Free-Free case scale
nicely with the size $N$ of the chain according to the dynamical
critical exponent $z$. To be more precise, let us introduce 
the fixed point values of those matrices denoted by 
$A^{\ast}$, $B^{\ast}_L$, $B^{\ast}_R$ and $C^{\ast}$ which we
define as,

\begin{equation}
A^{\ast} = N^{-z} a^{\ast}, \ \
B^{\ast}_L = N^{-z} b^{\ast}_L, \ \
B^{\ast}_R = N^{-z} b^{\ast}_R, \ \
C^{\ast} = N^{-z} c^{\ast},  \ \ \mbox{Fixed-Point values}\label{cb11f}
\end{equation}

\noindent in terms of the scaled matrices 
$a^{\ast}$, $b^{\ast}_L$, $b^{\ast}_R$ and $c^{\ast}$. For a block
of 3 sites ($n_s=3$) we find the following Fixed-Point structure
parametrized by two constants $s$ and $t$ (for bigger $n_s$ we 
need extra parameters),

\begin{equation} a^{\ast} =  0, \ 
 b^{\ast}_R =  \left( \begin{array}{ccc}  
1& s & s\\
 s& t &t \\
  s& t & t   \end{array}      \right)  , \  
b^{\ast}_L =  \left( \begin{array}{ccc}  
1&  -s& s\\
-s & t & -t\\
  s& -t & t
   \end{array}      \right)  , \  
c^{\ast} =  \left( \begin{array}{ccc}  
-1& s &-s \\ 
-s & t & -t\\
 -s & t & -t
   \end{array}      \right)   \label{cb11ff}
      \end{equation}

\noindent with $s=1.3993$ and $t=1.9581$. The critical exponent $z$
we obtain is,

\begin{equation} 
z = 0.9999 \label{cb11fff}
      \end{equation}

\noindent which is indeed very close to the exact value $z=1$ 
(actually, it differs in the ninth decimal digit).

\noindent The interpretation of this Fixed-Point in the context of
the CBRG method is as follows. We pointed out before that when the 
boundary operators $b_{L,R}$ vanish we recover the standard BRG 
method in which the blocks are not correlated. Here we find that it
is the uncorrelated Hamiltonian which vanish, while the boundary
$b_{L,R}$ and interaction $c$ operators do not vanish within the
scaling law.
This fact may perhaps be interpreted by saying that in the example
under study the correlation between blocks is more important
than their selfenergy. In references \cite{nishino},
\cite{ostlund-rommer}, \cite{nishino-okunishi} it was shown that the
DMRG method leads, in the thermodynamic limit, to a
``product form" ansatz for the
ground state wave function. In our case we see from Eqs.(\ref{cb11f}),
(\ref{cb11ff}) that we also reach thermodynamical limit, which leads us
to ask about the nature of the ansatz for the ground state and 
excited states implied by the CBRG method. The answer to this 
question will be addressed in a future publication but it suffices
to say that both the DMRG and the CBRG methods seem to yield 
different ansatzs of the ground state wave function.
In a few words, the DMRG is associated with a ``vertex picture"
while the CBRG is associated with a ``string picture".

\section{The Two-Dimensional CBRG-Algorithm}

The RG-method that we have devised in the one-dimensional problem
can be generalized in a natural way to higher dimensions. We shall
consider for simplicity the 2D case. First of all, we divide the
square lattice into blocks of $n_s$  sites each. Each block will in
turn be a square lattice with  a minimum of 4 sites ($= 2\times 2$
block). As in 1D, we shall define the following Hamiltonians to
carry out the CBRG-program,

\begin{itemize}

\item $A_p =$ self-energy of the $p$-th block isolated from the 
lattice.

\item $B_{p,q} =$ self-energy of the $p$-th block induced by the 
presence of the $q$-th block.

\item $C_{p,q} =$ interaction between the $p$-th block and the
$q$-th block.

\end{itemize}

\noindent The difference with respect to the 1D case is that each
block has now 4 neighbours  and therefore there are four different
$B$ and $C$ matrices.

Let us consider again the Hamiltonian of a free particle moving  in
a 2D-box with Free BC's at the boundaries of the box. The 2D
Hamiltonian is given again by the incidence matrix of the  lattice.
As in 1D we shall choose the matrix $A$ as the incidence matrix of
the block. Thus, for example, for a $2\times 2$ block  we have,

\begin{equation} A =  \left( \begin{array}{cccc}   2 & -1 & 0 & -1\\
-1 & 2 & -1 & 0\\ 0 & -1 & 2 & -1\\ -1 & 0 & -1 & 2
   \end{array}      \right) \label{cb12}
      \end{equation}

\noindent The 4 Boundary Operators $B$ 
have a  diagrammatic representation \cite{cbrg} which helps us
to keep track of their location in the block $H_B$ and interblock
$H_{BB}$ Hamiltonians.  Their explicit
matricial form is as follows,

\begin{equation} B_{12} = B_{43} = B_L =  \left(
\begin{array}{cccc}   0 &  &  & \\
 & 1 &  & \\
 &  & 1 & \\
 &  &  & 0
   \end{array}  \right), \ \  B_{21} = B_{34} = B_R =  \left(
\begin{array}{cccc}   1 &  &  & \\
 & 0 &  & \\
 &  & 0 & \\
 &  &  & 1
   \end{array}  \right) \label{cb13}
      \end{equation}

\begin{equation} B_{14} = B_{23} = B_D =  \left(
\begin{array}{cccc}   0 &  &  & \\
 & 0 &  & \\
 &  & 1 & \\
 &  &  & 1
   \end{array}  \right), \ \  B_{41} = B_{32} = B_U =  \left(
\begin{array}{cccc}   1 &  &  & \\
 & 1 &  & \\
 &  & 0 & \\
 &  &  & 0
   \end{array}  \right) \label{cb14}
      \end{equation}

\noindent where the labels denote the position of the neighbouring
blocks and we have used the translation invariance of the 2D
tight-binding model so that we need only to distinguish between
Right and Left, and Up and Down.

\noindent As for the Interaction $C$-Operators \cite{cbrg} 
we have the following matricial
representation, with the same considerations as for the
$B$-operators,

\begin{equation} C_{12} = C_{43} = C_{LR} =  \left(
\begin{array}{cccc}   0 & 0 & 0 & 0\\ -1 & 0 & 0 & 0\\
 0& 0 & 0 & -1\\
 0&  0& 0 & 0
   \end{array}  \right), \ \  C_{21} = C_{34} = C_{RL} =  \left(
\begin{array}{cccc}   0 & -1 & 0 & 0\\ 0 & 0 & 0 & 0\\
 0& 0 & 0 & 0\\
 0&  0& -1 & 0
   \end{array}  \right) \label{cb15}
      \end{equation}

\begin{equation} C_{14} = C_{23} = C_{DU} =  \left(
\begin{array}{cccc}   0 & 0 & 0 & 0\\ 0 & 0 & 0 & 0\\
 0& -1 & 0 & 0\\
 -1&  0& 0 & 0
   \end{array}  \right), \ \  C_{41} = C_{32} = C_{UD} =  \left(
\begin{array}{cccc}   0 & 0 & 0 & -1\\ 0 & 0 & -1 & 0\\
 0& 0 & 0 & 0\\
 0&  0& 0 & 0
   \end{array}  \right) \label{cb16}
      \end{equation}

\noindent Thus translation invariance reduces the number of
independent CBRG-matrices by a half. These relations are particular
of the problem at hand but we must left open the posibility of
having all those matrices different from each other in order to
handle more complicated problems.

\noindent Now that we have all the elements entering in our
CBRG-method we proceed to construct the block $H_{sB}$ and interblock
$H_{sB,sB}$
 Hamiltonians out of them. To this end we have to consider a
superblock made up of 4 blocks \cite{cbrg}. Thus, for $H_{sB}$ we have,

\begin{equation} H_{sB} =  \left( \begin{array}{cccc}   A + B_L +
B_D & C_{LR} & 0 & C_{DU}\\ 
C_{RL}  & A + B_R + B_D & C_{DU}& 0\\
 0& C_{UD} & A + B_R + B_U & C_{RL}\\
 C_{UD} &  0& C_{LR} & A + B_L + B_U
   \end{array}  \right) \label{cb17}
      \end{equation}

\noindent This is a $4 n_s \times 4 n_s$ matrix made up of $ n_s
\times  n_s$  matrices.

\noindent As for the interblock Hamiltonian $H_{sB,sB}$ we have to
distinguish between $(sB,sB)$-couplings of horizontal type denoted by
$H^{(hor)}_{sB,sB}$ and vertical type denoted by
$H^{(ver)}_{sB,sB}$, which read explicitly as, 

\begin{equation}
H^{(hor)}_{sB,sB} =  \left( \begin{array}{cccccccc}   B_R &  &  &  &
0 & C_{RL} & 0 & 0\\
 & 0 & &  & 0 & 0 & 0 & 0\\
 & & 0 &  & 0 & 0 & 0 & 0\\
 & & & B_R & 0 & 0 & C_{RL} & 0 \\ 0 & 0 & 0 & 0 & 0 &  &  & \\
 C_{LR}& 0 &0 & 0 &  & B_L& & \\
 0& 0& 0 & C_{LR} &  &  & B_L & \\
 0& 0& 0& 0 &  &  &  & 0 
   \end{array}  \right), \quad
H^{(ver)}_{sB,sB} =  \left(
\begin{array}{cccccccc}   B_D &  &  &  & 0 & 0 & 0 & C_{UD}\\
 & B_D & &  & 0 & 0 & C_{UD} & 0\\
 & & 0 &  & 0 & 0 & 0 & 0\\
 & & & 0 & 0 & 0 & 0 & 0 \\ 0 & 0 & 0 & 0 & 0 &  &  & \\
 0& 0 &0 & 0 &  & 0& & \\
 0& C_{DU}& 0 & 0 &  &  & B_U & \\
 C_{DU}& 0& 0& 0 &  &  &  & B_U
   \end{array}  \right) 
\label{cb18a}
      \end{equation}

\noindent where we have made use again of translational invariance.

\noindent Once that we have made our choice for the decomposition of
the total Hamiltonian of the 2D-tight-binding model into block and
interblock Hamiltonians according to our CBRG-prescription, we can
carry on with the truncation part of the RG-method. We shall keep
$n_s$ states out of $4 n_s$ states per superblock so that our
truncation scheme may be summarized as,

\[ 4 n_s \ \mbox{(superblock)} \longrightarrow n_s \ \mbox{(new
block)} \]

Recall that at each step of the CBRG-method we need to identify the 
$A$, $B_L$, $B_R$ and $C$ operators which define the truncation
procedure for  the next step of the method. For this purpose,
firstly the truncation of the  superblock $H_{sB}$ gives rise to the
$A'$ uncorrelated self-energy operator for the next RG-step, namely,

\begin{equation} H_{sB}  \  \mbox{( $4 n_s \times 4 n_s$
matrix)}\longrightarrow A' \  \mbox{($n_s \times n_s$ matrix)}
\label{cb19}
      \end{equation}

\noindent To identify the rest of the operators we have to
renormalize the interblock Hamiltonian which comes in two types,
horizontal and vertical. The
renormalization of the $H^{(hor)}_{sB,sB}$ Hamiltonian is given by
\cite{cbrg},

\begin{equation} H^{(hor)}_{sB,sB} \longrightarrow  \left( \begin{array}{cc}  
B'_R & C'_{RL}  \\
 C'_{LR}& B'_L 
   \end{array}  \right) \label{cb20}
      \end{equation}

\noindent Likewise, for the $H^{(ver)}_{sB,sB}$ Hamiltonian we have,

\begin{equation} H^{(ver)}_{sB,sB} \longrightarrow  \left( \begin{array}{cc}  
B'_D & C'_{UD}  \\
 C'_{DU}& B'_U 
   \end{array}  \right) \label{cb21}
      \end{equation}

\noindent Now that we have identified all the operators defining the
CBRG method at the new stage of the renormalization, we may
reconstruct the new superblock Hamiltonian $H'_{sB}$, which in turn
has the same form as the original  $H_{sB}$ in Eq.(\ref{cb17})
substituting all the operators by their {\em primed versions}.
This statement can be explicitly checked by considering the 
set of 4 superblocks \cite{cbrg}.
Firstly, the new $H'_{sB}$ has a contribution coming from the 
truncation of each of the 4 superblocks, each of them 
contributing with an $A'$-operator as in Eq.(\ref{cb19}).
Secondly, $H'_{sB}$ picks up two more contributions coming from
the horizontal and vertical interaction between superblocks,
which we denote by $H_{\leftrightarrow}$ and $H_{\updownarrow}$.
Thus, in the CBRG-method $H'_{sB}$ is renormalized as,

\[
H'_{sB} =  \left( \begin{array}{cccc}   
A'  &  &  & \\ 
  & A'  & & \\
 & & A' & \\
  &  &  & A'
   \end{array}  \right) \ \ \longleftarrow 
\mbox{(single superblock contribution)}
\]
\[ 
(H_{\leftrightarrow}) \ \rightarrow \ + 
\left( \begin{array}{cccc}   
B'_L  & C'_{LR} &  & \\ 
C'_{RL}  & B'_R  & & \\
 & & 0 & \\
  &  &  & 0
   \end{array}  \right) + 
\left( \begin{array}{cccc}   
0  &  &  & \\ 
  & 0  & & \\
 & & B'_R &C'_{RL} \\
  &  & C'_{LR} & B'_L
   \end{array}  \right)
\]
\begin{equation} 
(H_{\updownarrow}) \ \rightarrow \ + 
\left( \begin{array}{cccc}   
B'_U  &  &  & C'_{DU}\\ 
  &  0 & & \\
 & & 0 & \\
 C'_{UD} &  &  & B'_D
   \end{array}  \right) + 
\left( \begin{array}{cccc}   
0  &  &  & \\ 
  & B'_U  & C'_{DU} & \\
  & C'_{UD} &B'_D & \\
  &  &  & 0
   \end{array}  \right)
\label{cb23}
      \end{equation}

\noindent and altogether we arrive at the previously stated result
of Eq.(\ref{cb17}).

\noindent Similarly we may proceed with the renormalized
interblock Hamiltonians $\mbox{$H'$}^{(hor)}_{sB,sB}$ (\ref{cb20}) and
$\mbox{$H'$}^{(ver)}_{sB,sB}$ (\ref{cb21})
and we end up with the same form for them as the original ones.

\noindent This ends the implementation of the CBRG-method for the 
2D-tight-binding model.

In Table 5 we collect our CBRG results for the first 4 lowest lying
states for a chain of $N=4\times 4\times 4^6=65536$ sites. Comparison
with the exact results gives a good agreement. We have also data
from truncations with blocks of $n_s=9$ and $n_s=16$ sites which 
enforce this statement.
Moreover, notice that the first excited state is a doublet as in the
exact solution.

Another important result of our CBRG-method is that the 
$(n^2_1 + n^2_2)/N^2$ scaling law for the energy of the 
$(n_1,n_2)$-excited states of a square lattice of length $N$ is 
reproduced correctly. In fact, from data of the $n_s=16$ sites
truncation for the first-excited-state energy we can extract the
corresponding $1/N^2$-scaling law which turns out to be,

\begin{equation}
 E^{(CBRG)}_1 (N) = c^{(1)}_{CBRG} \frac{1}{N^2}, \ \ 
c^{(1)}_{CBRG} = 9.7365, 
 \ (N \longrightarrow \infty) \ \mbox{D=2 Free BC's}\label{cb24} 
  \end{equation}

\noindent while the exact value of the proportionality constant
$c$ is $c_{exact}=\pi^2=9.86$. This amounts to a 1.3 \% error.

Likewise, we may obtain the full $(n^2_1 + n^2_2)/N^2$ scaling law
for the whole set of 15 excited states and we find,

\begin{equation}
 E^{(CBRG)}_{(n_1,n_2)} (N) = c^{(1)}_{CBRG} 
\frac{(n^2_1 + n^2_2)}{N^2}, \ \ 
c_{CBRG} = 7.9074, 
 \ (N \longrightarrow \infty) \ \mbox{D=2 Free BC's}\label{cb25} 
  \end{equation}

\noindent which now amounts to a 10.5 \% error.

As in the 1D Free-Free case, we can determine critical scaling 
exponent $\theta$ (\ref{cb8abc}). For a truncation scheme 
$16 \rightarrow 4$ we find,

\begin{equation}
 \theta = 1.99999981 \ \ \ \mbox{D=2}\label{cb26} 
  \end{equation}

\noindent which clearly supports the scaling laws introduced above.
Notice again (see Table 5) that our CBRG method gives the exact
(within machine precision) energy of the ground state. This is true
for every step of the RG, as was proved in \cite{bc-germanyo} for 1D.

We can also perform the wave function reconstruction of the excited
states in the two-dimensional real space. This is achieved by a 
two-dimensional extension of the reconstruction equation (\ref{cb9}).
As an illustration of how the CBRG method performs with this matter,
see \cite{cbrg}.
The qualitative real-space form of the excited-state wave 
functions are captured by the CBRG procedure.

\noindent With this discussion we close the first part of these notes which have
been devoted to new develoments of the Real-Space
 RG method revolving around the new
ideas brought about by the Density Matrix RG method.


\section{Standard BRG for the 1D AF Heisenberg Model}

In the remaining sections we shall return to the standard BRG methods to deal
in an analytical controlled fashion with models which include many-body interactions
unlike the free models considered in the first part of these notes.

\noindent The arquetypical model we shall be dealing with is the Heisenberg model
which will be studied in ladder systems and 2D lattices (square, honeycomb).
Our point will be that even with the old-fashion BRG
can be useful to retrieve the correct qualitative phyisics when properly implemented.
To this end we shall be needing some results concerning the Heisenberg model
in one dimension which will be basic.

Let us recall  the standard BRG method for 
the AF Heisenberg-Ising model whose Hamiltonian is given by:

\begin{equation}
H_N = J \sum _{j=1}^{N-1} (S^x_j S^x_{j+1} + S^y_j S^y_{j+1} + \Delta S^z_j S^z_{j+1})   \label{h9}
\end{equation}

\noindent where $\Delta \geq 0$ is the anisotropic parameter and $J>0$ for the antiferromagnetic 
case. If $\Delta = 1$ one has the AF-Heisenberg model which was solved by Bethe in 1931. If 
$\Delta = 0$ one has the XX-model which can be trivially solved using a Jordan-Wigner transformation
which maps it onto a free fermion model. For the remaing values of $\Delta $ the model is also 
solvable by Bethe ansatz and it is the 1D relative of the 2D statistical mechanical model known as 
 the 6-vertex or XXZ-model.  

\noindent The region $\Delta > 1$ is massive with a doubly degenerate ground state 
in the thermodynamic limit $N \rightarrow \infty$ characterized 
by the non-zero value of the staggered magnetization,

\begin{equation}
m_{\mbox{st}} = \langle \frac{1}{N} \sum_j S^z_j (-1)^j  \rangle   \label{h9b}
\end{equation}

\noindent The region $0 \leq \Delta \leq 1$ is massless and the ground state is non-degenerate 
with a zero staggered magnetization. The phase transition between the two phases has an 
essential singularity.

We would like next to show  which of these features are captured by a real-space RG-analysis.
The rule of thumb for the RG-approach to half-integer spin model or fermion model is to consider 
blocks with an {\em odd number of sites}. This allows in principle, although not necessarilly, to 
obtain effective Hamiltonians with the same form as the original ones. Choosing for (\ref{h9}) blocks 
of 3 sites we obtain the block Hamiltonian:

\[
\frac{1}{J} H = \vec{S}_1 \cdot \vec{S}_2 +   \vec{S}_2 \cdot \vec{S}_3 + 
\epsilon (S^z_1 S^z_2 + S^z_2 S^z_3)  
\]
\begin{equation}
= \frac{1}{2} \left\{ [\vec{S}_1 + \vec{S}_2 +\vec{S}_3]^2 -
 (\vec{S}_1 + \vec{S}_3)^2 - 3/4 \right\}   + \epsilon (S^z_1 S^z_2 + S^z_2 S^z_3) 
 \label{h10}
\end{equation}

\noindent $\epsilon := \Delta - 1$.

If $\epsilon = 0$ the block Hamiltonian $H_B$ is invariant under the $SU(2)$ group and according 
to the introduction to this section,  we should consider the tensor product decomposition:

\begin{equation}
\frac{1}{2} \otimes \frac{1}{2} \otimes \frac{1}{2}  = \frac{1}{2}  \oplus \frac{1}{2} 
\oplus \frac{3}{2}                                                                                       \label{h11}
\end{equation}

\noindent The particular way of writing $H_B$ given in Eq. (\ref{h10}) suggests to compose first 
$\vec{S}_1$ and  $\vec{S}_3$ and then, the resulting spin with $\vec{S}_2$. The result of this 
of this compositions is given as follows:

\begin{equation}
| \frac{3}{2},  \frac{3}{2} \rangle = | \uparrow \uparrow \uparrow \rangle  \ \ E_B = J/2     
  \label{h12a}
\end{equation}

\begin{equation}
| \frac{3}{2},  \frac{1}{2} \rangle = \frac{1}{\sqrt{3}} 
( | \uparrow \downarrow \uparrow \rangle  +  | \downarrow \uparrow \uparrow \rangle  +      
   | \uparrow \uparrow \downarrow \rangle  )     \ \ E_B = J/2             \label{h12b}
\end{equation}

\begin{equation}
| \frac{1}{2},  \frac{1}{2} \rangle_1 = \frac{1}{\sqrt{2}} 
 (| \uparrow \uparrow \uparrow \rangle - | \downarrow \uparrow \uparrow \rangle) 
\ \ \ E_B = 0                                                                                         \label{h12c}
\end{equation}

\begin{equation}
| \frac{1}{2},  \frac{1}{2} \rangle_0 = \frac{1}{\sqrt{6}} 
( 2 | \uparrow \downarrow \uparrow \rangle  -  | \downarrow \uparrow \uparrow \rangle  -      
   | \uparrow \uparrow \downarrow \rangle  )     \ \ E_B = -J             \label{h12d}
\end{equation}

\noindent Hence for $\Delta =0$ we could choose the spin $1/2$ irrep. with basis vectors 
$|\frac{1}{2}, \frac{1}{2} \rangle_0$ and $|\frac{1}{2}, -\frac{1}{2} \rangle_0$ in order to define the 
intertwiner operator $T_0$.

\noindent However, if $\Delta \neq 0$ the states (\ref{h12a}) -(\ref{h12d}) are not eigenstates of 
(\ref{h10}). The full rotation group is broken down to the rotation around the z-axis. The states 
  $| \frac{3}{2},  \frac{1}{2} \rangle$ and $| \frac{1}{2},  \frac{1}{2} \rangle_1 $ are mixed in the new ground
state which is given by:

\begin{equation}
| + \frac{1}{2} \rangle = \frac{1}{\sqrt{1+2 x^2}} 
( 2 | \frac{1}{2},  \frac{1}{2} \rangle_1 + \sqrt{2} x| \frac{3}{2},  \frac{1}{2} \rangle ) \label{h13a}
\end{equation}

\noindent where 

\begin{equation}
x = \frac{ 2 (\Delta -1)}{8 + \Delta + 3 \sqrt{\Delta^2 + 8}}              \label{h13b}
\end{equation}

\noindent and its energy is,

\begin{equation}
E_B = -\frac{J}{4} [\Delta + \sqrt{\Delta ^2 + 8}]                                       \label{h13c}
\end{equation}

\noindent along with its $| - \frac{1}{2} \rangle$ partner. This are now the two states retained in the 
RG method. To be more explicit, we have

\begin{equation}
| + \frac{1}{2} \rangle = \frac{1}{\sqrt{6(1+2 x^2)}} 
[(2 x + 2) |\uparrow \downarrow \uparrow \rangle + 
(2 x - 1) |\uparrow \uparrow \downarrow \rangle + 
(2 x - 1) |\downarrow \uparrow \uparrow \rangle ] \label{h14a}
\end{equation}

\begin{equation}
| - \frac{1}{2} \rangle = -\frac{1}{\sqrt{6(1+2 x^2)}} 
[(2 x + 2) |\downarrow \uparrow \downarrow \rangle + 
(2 x - 1) |\downarrow \downarrow \uparrow \rangle + 
(2 x - 1) |\uparrow \downarrow \downarrow \rangle ]                              \label{h14b}
\end{equation}

\noindent The intertwiner operator $T_0$ reads then,

\begin{equation}
T_0 = | + \frac{1}{2}\rangle \langle ' \uparrow | +    
| - \frac{1}{2}\rangle \langle ' \downarrow |                                   \label{h15}
\end{equation}

\noindent where $|\uparrow \rangle'$  and $|\downarrow \rangle'$ form a basis for 
the space 
$V'={\bf C}^2$. The RG-equations for the spin operators $\vec{S}_i$ ($i=1,3$) are 
then given by 

\begin{equation}
T_0^{\dag} \vec{S}^x_i T_0 =    \xi^x \vec{S' }^x_i   \ \ i = 1,3.                   \label{hh15a}
\end{equation}

\begin{equation}
T_0^{\dag} \vec{S}^y_i T_0 =    \xi^y  \vec{S' }^y_i     \ \ i = 1,3.                   \label{hh15b}
\end{equation}

\begin{equation}
T_0^{\dag} \vec{S}^z_i T_0 =    \xi^z  \vec{S' }^z_i     \ \ i = 1,3.                   \label{hh15c}
\end{equation}

\noindent where $\xi^x$, etc are the renormalization factors which depend upon the anisotropy 
parameter by,

\begin{equation}
\xi^x = \xi^y := \frac{2 (1 + x) (1 -2 x)}{3 (1 + 2 x^2)}          \label{hh15d}
\end{equation}

 \begin{equation}
\xi^z  := \frac{2 (1 + x)^2}{3 (1 + 2 x^2)}          \label{hh15e}
\end{equation}

\noindent Observe the symmetry between the sites $i=1$ and $3$ which is a consequence of the
even parity of the states (\ref{h14a}) -(\ref{h14b}).

The renormalized Hamiltonian can be easily obtained using Eqs.(\ref{hh15a})-(\ref{hh15e}) 
 and (\ref{h9}), and 
apart from and additive constant it has the same form as $H$, namely,

\begin{equation}
T_0^{\dag} H_N (J,\Delta )T_0 =    \frac{N}{3} e_B (J,\Delta ) + 
         H_{N/3} (J',\Delta ')                                                                   \label{h16}
\end{equation}

\noindent where 

 \begin{equation}
J' = (\xi ^x)^2 J                                                                     \label{h17a}
\end{equation}

 \begin{equation}
\Delta' = (\frac{\xi ^z}{\xi^x})^2 \Delta                                                                     \label{h17b}
\end{equation}

\noindent Iterating these equations we generate a family of Hamiltonians 
$H^{(m)}_{N/3^m} (J^(m),\Delta ^{(m)})$. The energy density of the ground state of $H_N$ in the 
limit $N\rightarrow \infty $ is then given by,

 \begin{equation}
\lim _{N\rightarrow \infty} \frac{E_0}{N} = e^{BRG}_{\infty} = 
\sum _{m=0}^{\infty} \frac{1}{3^{m+1}} e_B (J^{(m)},\Delta^{(m)})          \label{h18}
\end{equation}

\noindent where initially $J^{(0)}=J$, $\Delta^{(0)}=\Delta$ and Eqs.(\ref{h17a}) -(\ref{h17b}) 
provide the flow of the coupling constants.

The analysis of Eq.(\ref{h17b}) shows that there are 3 fixed points corresponding to the values 
$\Delta =0$ (isotropic XX-model), $\Delta =1$ (isotropic Heisenberg model) and $\Delta =\infty$ 
(Ising model). The properties of these fixed points are given in Table 6.

\noindent The computation of $e^{BRG}_{\infty}$ in this case is facilitated by the fact that (\ref{h18}) 
becomes a geometric series at the fixed point. The exact results concerning the models $\Delta=0$ 
and $\Delta =1$ are extracted from references \cite{lieb-s-m} and \cite{orbach}. The case with 
$\Delta \rightarrow \infty$ is exact because the states $|\pm \frac{1}{2} \rangle$ given in 
(\ref{h14a}) - (\ref{h14b})
tend in that limit to the exact ground state $|\uparrow \downarrow \uparrow \rangle$ and 
$|\downarrow \uparrow \downarrow \rangle$ of the Ising model. As a matter of fact,

\[ 
|+\frac{1}{2} \rangle \simeq _{\Delta \rightarrow \infty} |\uparrow \downarrow \uparrow \rangle - 
\frac{1}{\Delta} |\uparrow \uparrow \downarrow \rangle - 
\frac{1}{\Delta} |\downarrow \uparrow \uparrow \rangle
\]

 \[ 
|-\frac{1}{2} \rangle \simeq _{\Delta \rightarrow \infty} -|\downarrow \uparrow \downarrow \rangle - 
\frac{1}{\Delta} |\downarrow \downarrow \uparrow \rangle - 
\frac{1}{\Delta} |\uparrow \downarrow \downarrow \rangle
\]

The region $0 < \Delta < 1 $ which flows under the RG-transformation to the XX-model is massless 
since both $J^{(m)}$ and $\Delta ^{(m)}$ go to zero. We showed at the begining of this section 
that all this region is critical (a line of fixed points) and therefore massless. The RG-equations 
(\ref{h17a}) -(\ref{h17b}) are not able to detect this criticality except at the point $\Delta =0$. Only 
the masslessness property is detected.

The region $\Delta >1$ which flows to the Ising model is massive and this follows from the fact 
that the product $J^{(m)} \Delta ^{(m)}$ goes in the limit $m \rightarrow \infty$ to a constant quantity
$J^{(\infty)} \Delta ^{(\infty)}$ which can be computed from Eqs. (\ref{h17a}) -(\ref{h17b}) and 
(\ref{h15}),

 \begin{equation}
  J^{(\infty)} \Delta ^{(\infty)} = \prod _{m=0}^{\infty} 
\frac{4}{9}  \frac{(1 + x_m)^4}{(1 + 2 x_m^2)^2}   \label{h19}
\end{equation}

\noindent where $x_m$ is given by (\ref{h13b}) with $\Delta $ replaced $\Delta ^{(m)}$. This quantity 
gives essentially the mass gap above the ground state and also the end-to-end or LRO order (Long 
Range Order) given by the expectation value $|\langle \vec{S}(1) \cdot \vec{S}(N) \rangle|$ in the limit 
$N \rightarrow \infty$.

In summary, the properties of the Heisenberg-Ising model are qualitatively and quantitatively well 
described in the massive region $\Delta >1$ while in the massless region $0 < \Delta < 1$ one 
predicts the massless spectrum {\em but no criticality at each value of $\Delta $}. This latter 
fact is rather subtle and elusive. One would like to construct a RG-formalism such that the 
Hamiltonian $H_N(\Delta )$ would be a fixed point Hamiltonian for every value of $\Delta $ in the 
range from -1 to 1. 

  The phase transition between the two regimes is correctly predicted to happen  at the value 
$\Delta = 1$. This is a consequence of the rotational symmetry, namely at $\Delta = 1$ the 
system is $SU(2)$ invariant and the RG transformation has been defined as to preserve this 
symmetry. When $\Delta \neq 1$ the $SU(2)$ symmetry is broken and this is reflected later on 
in the RG-flow of the coupling constant $\Delta $.


\section{RG for Heisenberg Spin Ladders}

Much of the El Escorial Summer School has been devoted to the nowdays 
very active field known as ladders systems (spin, t-J, Hubbard ...), see
 \cite{schulznote}, \cite{sierranote}, \cite{whitenote}.
 What does the BRG method have to say
on these systems? We again emphasize that this is a technically simple method
which produces qualitative correct results when properly applied.
Later it is possible to look for numerical accuracy using DMRG, second order
RG (see appendix) or some other means.

The Hamiltonian of a Heisenberg spin ladder with $n_l$ legs, each of length $N$ is
given by,

\begin{eqnarray}
   {\cal H}_{ladder} = {\cal H}_{leg} + {\cal H}_{rung} \cr
   {\cal H}_{leg} =  \sum_{a=1}^{n_l} \; \sum_{n=1}^{N} \; 
J
   {\bf S}_{a}(n) \cdot {\bf S}_{a}(n+1) \cr
 {\cal H}_{rung} =  \sum_{a=1}^{n_l-1} \; \sum_{n=1}^{N} \; 
J^\prime \;   {\bf S}_{a}(n) \cdot {\bf S}_{a}(n+1) 
\label{sl1}
\end{eqnarray}
\noindent
where ${\bf S}_{a}(n)$ are spin-$1/2$ matrices acting on the $a$-th leg at the position
$n=1,\ldots,N$, and $J$ is the intraleg coupling constant while 
$J^\prime$ is the interleg exchange coupling constants, both beign positive to 
guarantee AF spin ladders.

We shall concentrate on the uniform Heisenberg ladders with no staggering.
There are many examples that can be worked out but to be concrete we shall
pick up the 3-leg ladder system \cite{gyunp}. 
In advance, what we are going to obtain is the RG-flow towards
the strong coupling limit of spin ladders.
This is an alternative to the determination of the RG-fow using 
bosonization as performed by H.J. Schulz \cite{prb86}, \cite{schulznote}.
To apply the BRG we need to set up what is
the block Hamiltonian $H_B$ which we do by forming blocks of 3 sites each
along every leg and located one block on top of another as in fig.7
In this fashion we are selecting a subset of couplings from the whole spin ladder
Hamiltonian in (\ref{sl1}). The remaining terms involving links with neighbouring
blocks make up for the interblock Hamiltonian $H_{BB}$. We shall not write down
explicitely the analytical expressions for $H_B$ and $H_{BB}$ as it is quite clear
what is meant simply by looking at fig.7.

\noindent Now it is aparent that the standard results of the previous section are at
work for spin ladders. Notice that $H_B$ is made up of small block Hamiltonians
of 3 sites as in (\ref{h10}) whose eigenstates and energies are already computed in
(\ref{h12a})-(\ref{h12d}). The renormalization process goes through all the way
by truncating the block states to the lowest eigenstates, i.e., the spin 
doublet $|{1\over 2}, \pm {1\over 2}\rangle_0$, with energy $e_0=-J$. In this case
the embedding operator $T^{(\alpha)}$ 
for each block is nothing but the projector $P_0^{(\alpha)}$ onto these
states; denoting

\begin{equation}
P_B = \prod_{\alpha =1}^{N/3} P_0^{(\alpha)}
\label{sl2}
\end{equation}

\noindent then, to first order
in $J$ the renormalized or effective Hamiltonian acting on the states left out after the
truncation is simply,

\begin{equation}
H_{\text{eff}} = P_B (H_B + H_{BB}) P_B\label{sl4b}
\end{equation}

\noindent Using the embedding operator (\ref{h15}) we arrive at,

\begin{equation}
P_B H_B  P_B = + {N\over 3} e_0 = -{N\over 3} J \label{sl5b}
\end{equation}

\noindent In this case the renormalization of the block Hamiltonian gives the identity 
because the two states retained within each block are degenerate by rotational invariance.
The renormalization of the interblock Hamiltonian is also simple if we observe that 
$H_{BB}$ contains products of spin operators belonging to different blocks, say 
${\bf S}_i^{\alpha} \cdot {\bf S}_j^{\beta}$ where $\alpha \neq \beta$ denote neighbouring 
blocks and $i,j = 1,2,3$ are the intrablock labels used in Eq. (\ref{sl1}). Then, according
to Eq. (\ref{sl2}) we have,

\begin{equation}
P_B {\bf S}_i^{\alpha} \cdot {\bf S}_j^{\beta} P_B = 
P_B (P_0^{(\alpha)}{\bf S}_i^{\alpha} P_0^{(\alpha)}) 
(P_0^{(\beta)}{\bf S}_i^{\beta} P_0^{(\beta)})  P_B\label{sl6bb}
\end{equation}

\noindent Hence we only need to know how the spin operators renormalize within each
block onto the new spin operators. By symmetry arguments, the renormalization spin
factor denoted by $\xi _i$ must be the same for the 3 components of the spin operators,
i.e.,

\begin{equation}
(P_0^{(\alpha)}{\bf S}_i^{\alpha} P_0^{(\alpha)}) = \xi _i {\bf S}^\prime _{\alpha}
\label{sl7b}
\end{equation}

\noindent where ${\bf S}^\prime _{\alpha}$ denotes the spin 1/2 operator acting on the
effective spin 1/2 subspace of the $\alpha^{\text{th}}$-block. The renormalization 
spin factors are known from (\ref{hh15d})-(\ref{hh15e}) to be given by,

\begin{eqnarray}
\xi_1 = \xi_3 = {2\over 3} \cr
\xi_2 = - {1\over 3}
\label{sl9b}
\end{eqnarray}

Now we are ready to compute the renormalization of the interblock Hamiltonian
$H_{BB}$. According to Eqs. (\ref{sl6bb}) and (\ref{sl7b}), the renormalization of the
horizontal couplings between blocks of $H_{BB}$ (see Fig. 7) is given by,

\begin{equation}
J \; {\bf S}_3^{(\alpha)} \cdot {\bf S}_1^{(\beta)}  \longrightarrow 
J \; \xi_1 \xi_3 \; {\bf S}_a (n^\prime) {\bf S}_a (n^\prime+1) = {4\over 9} J \;
 {\bf S}_a (n^\prime) {\bf S}_a (n^\prime+1)
\label{sl11b}
\end{equation}

\noindent while the 3 vertical couplings between two blocks are renormalized to

\begin{eqnarray}
J^\prime \; ({\bf S}_1^{(\alpha)} \cdot {\bf S}_1^{(\beta)} + 
{\bf S}_2^{(\alpha)}\cdot {\bf S}_2^{(\beta)} + 
{\bf S}_3^{(\alpha)} \cdot {\bf S}_3^{(\beta)})  \longrightarrow 
J^\prime \; (\xi_1^2 + \xi_2^2 + \xi_3^2) \; {\bf S}^{(\alpha)} \cdot {\bf S}^{(\beta)} \cr
= J^\prime {\bf S}_a (n^\prime) \cdot {\bf S}_{a+1} (n^\prime+1) 
\label{12b}
\end{eqnarray}

\noindent We have then obtained that the renormalized Hamiltonian (\ref{sl4b}),
apart from the constant term (\ref{sl5b}), is the as the original ladder Hamiltonian,
but with length $N/3$ and the following renormalization of the coupling constants,

\begin{eqnarray}
J \longrightarrow {4\over 9} J \cr
J^\prime \longrightarrow J^\prime
\label{13b}
\end{eqnarray}

\noindent Hence the ratio $J^\prime / J$ increases as,

\begin{equation}
{J^\prime \over J} \longrightarrow {9\over 4} {J^\prime \over J} \label{14b}
\end{equation}

\noindent after each step of the RG showing that $J^\prime /J = \infty$ is a 
stable fixed point which controls the behaviour for all values of $J$ and $J^\prime$,
while $J^\prime /J = 0$ is an unstable fixed point. Had we chosen blocks made up
of more than 3 sites we would have obtained essentially the same result. 
The RG method for the simple spin chain (i.e. $n_l=1$) where first obtained in 
reference \cite{rabin}. 
\noindent According to Eq. (\ref{13b}) if we start in the weak coupling regime
$J^\prime /J \ll 1$, after sufficient iterations of the RG we would get an effective
Hamiltonian in the strong coupling regime where we can apply the arguments of
section II to derive the nature of the low lying spectrum of the theory.


\section{Real-Space RG Approach to the Quantum 2D-AF Heisenberg Model}

In this section we present a real-space RG treatment of the quantum two-dimensional
Heisenberg antiferromagnetic model with arbitrary spin $S$. Most of the work using
real-space RG methods has been devoted to one-dimensional problems. This is
very useful because it is crucial to have an aproximate method which gives good
results for both 1D and 2D problems, for it is known that mean field theory methods
fail in low dimensional problems. In this regard we have recently shown that the use
of quantum groups in combination of real-space methods in 1D captures the
essential features exhibited by the exact solutions of models such as Heisenberg 
and ITF \cite{q-germanyo}, \cite{qbis-germanyo}.
Nevertheless, the main reason which has prevented the applications
 of the real-space RG in 2D 
quantum lattice Hamiltonians is the rapid growth of the number of states to be
kept in a reasonable scheme of truncation of states in dimensions higher than one.

We shall be using the Block RG method in our study of the 2D Heisenberg model.
This version of the RG method is suitable to achieve fully analytical treatments of
interacting many-body problems. The reason for searching for complete analytical
approaches as opposed to purely numerical studies relies on the necessity of having
a qualitative understanding of the mechanisms responsible for the different behaviors
exhibited by the Heisenberg model itself and for its  connections
to more complicated related Hamiltonians such as t-J and Hubbard where the
understanding of the doping effects is a big issue at stake.
In order for there to be a completely analytical  RG treatment in 2D we need a 
juidicious choice of the states to be kept as we shall see \cite{rotating}.

Despite of some initial controversies there is by now sufficient
theoretical and experimental evidence for the existence of
antiferromagnetic long range order 
(AF LRO) in the 2d spin 1/2 Heisenberg antiferromagnet
\cite{manousakis} (and references therein). 
This property has been observed in parent
compounds of hight-${T_c}$ materials such as $ La_2 Cu O_4$
\cite{manousakis}. 
From a theoretical point of view this  means that the
strong quantum fluctuations implied by the low dimensionality
and low spin  do not destroy completely the Neel order, as it
happens \cite{bethe} in 1d. Though there is no  a satisfactory
physical explanation of this fact, which may be important regarding
 the interplay between  antiferromagnetism and superconductivity
upon doping. The RVB scenario originally proposed by Anderson
\cite{anderson,kivelson}, 
while yielding an appealing picture of the ground
state, does not explain the  presence of AF LRO. This type of
order may however be incorporated a posteriori in long range RVB
ansatzs of factorized form  \cite{liang-doucot-anderson}, 
with predictions
similar to the ones obtained using Quantum Monte Carlo methods
\cite{carlson} and variational plus Lanczos techniques 
\cite{hebb-rice}. 
A class of physical systems  where the RVB approach may
be actually realized is in spin ladders with an even number of
chains \cite{white-noack-scalapino,dagotto-rice}. The previous works
leave still room to investigate in more depth the interplay between
the RVB scenario, or more generally  ``valence bond scenarios'',
and the AF order present in the 2d AF-magnets, described by the
AF Hamiltonian 
$H=J \sum _{\langle i,j \rangle} {\bf S}_i \cdot {\bf S}_j$.

We have proposed a new scenario where the valence bonds,
instead of resonating as in the RVB scenario, rotate around their ends
under the influence of the AF background. To test this idea
we propose a variational ground state in which the bonds rotate but do
not resonate among themselves.
We shall start by considering how the quantum fluctuations affect the 
classical Neel state. This is also the starting point of the spin wave theory
(SW), which we would like to use for comparison of our theory.
An important ingredient of our construction is the use of real-space RG techniques,
which allows us to obtain exact analytical results for any value of the spin S
of the model (S is integer or half-integer and in the discussion above S=1/2.)
The advantage of using a real-space RG method is that one can treat in an
exact manner the local quantum fluctuations of the classical Neel state.
By this we mean that if we divide the square lattice into blocks of 5 sites each,
as in Figs. 8,9 and 10,
 then the Heisenberg Hamiltonian restricted to the blocks
can be solved exactly.
The ground state for every block is a spin 3/2 irrep. (if S=1/2) which is obtained
by forming a singlet (bond) between the spin at the center and the ones surrounding
the center (Fig. 8)
According to the RG method, the spin 3/2 can be chosen as an effective
spin for the renormalized lattice which now has N/5 sites. We shall show  later 
on that the interaction between those effective spins 3/2 (or 3S more generally) 
is also governed by an AF-Heisenberg model with a renormalized coupling constant.
Hence the RG procedure can be iterated yielding a series of effective spins
which ultimately goes to its classical value, i.e. infinity! ($S \rightarrow 3S 
\rightarrow 3^2S \rightarrow \infty$.)
We thus obtain in an economical and simple way the important result that
the 2D AF-Heisenberg models belongs to the universality class of the 2D classical
Heisenberg model. For a sigma model derivation of this result see ref. \cite{CHN}.
The rotating bond picture puts in correspondence various approaches to the
2D AF-Heisenberg model.

Let us begin our approach by considering the cluster of 5 spins 1/2
of Fig.8 a). The configuration showed in Fig.8 a)  is the exact
ground state of the Ising piece  of the Heisenberg Hamiltonian,
given by $H_{z} = J\sum_{i=1,\dots,4} S^z_0 S^z_i $ , where $S_0^z $
and $S_i^z$ are the third component of the spin operators at the
center and the $i^{th}$ position off the center respectively. 
As soon as the ``transverse'' Hamiltonian
 $H_{xy} = J \sum_{i=1,\dots,4} (S^x_0 S^x_i + S^y_0 S^y_i) $ is switched
on, the down-spin in the middle starts to move around the
cluster, and a valence bond between the center and the remaining
sites is formed in a s-wave ($l=0$) symmetric state as 
shown in Fig.8 b). Other rotational states with $l\neq 0$ may appear
corresponding to excitations ($l$ being the orbital angula momentum 
of the bond).
An alternative
description of this state is given by first combining the 4
spins sourrounding the center into a spin 2 irrep, which in turn is
combined with the spin 1/2 at the center yielding a spin 3/2 irrep
with energy $e_0 = - 3 J/2$. If instead of the spin 1/2 at
each site there is a spin S the previous analysis can be easily
generalized as follows: the ground state of the AFH Hamiltonian of
the 5-cluster has total spin 3S and is obtained by first combining
all the surrounding spins into a spin 4S, which in  turn
becomes a spin 3S after multiplication with the spin S at the
center. In a certain sense this state can be viewed as the formation
of bonds between the center and its four neighbours.
After applying several steps of the real-space RG, as we shall see 
below, new bonds are generated between sites at longer distances
apart. Thus our valence-bond scenario is a type of long range valence
bond state.

To study the AFH model in the entire square lattice we begin by first
tesselating this plane using the cluster of Fig.8 as the 
fundamental cell (see Fig. 9). Notice that the centers of the
5-cluster form a new square lattice with lattice spacing $a' =
\sqrt{5} a$.  Given this tesselation we can apply the standard RG
method of replacing clusters of spins by an effective spin
\cite{jullienlibro,jaitisi}. This method has been applied 
for the 1d AFH model by Rabin \cite{rabin} for clusters or blocks
with 3 sites, obtaining a ground state energy with an error of
$12\%$. 
The effective spin of every 3-block in 1d  has spin 1/2.
In our case, as we have discussed above, the effective spin of
the 5-blocks have spin 3S and the energy per block equal to $e_0
= - J S( 4S +1)$. The effective spins $S^\prime=3S$ interact by
means of an effective Hamitonian which to first order in
perturbation theory can be derived if we know the renormalization
of the spin operators ${\bf S}_{\alpha} \rightarrow \xi_{\alpha} 
{\bf S}^\prime ,
\alpha =  0, 1, \dots, 4$.

 The {\em renormalization spin
factor}  $\xi_{\alpha}$ can be shown to be given by the sum
$\xi_{\alpha}=\case{1}{3S} \; \sum_{m_0,m_1,\ldots,m_4} m_{\alpha}
(C^{3S}_{m_0,m_1,\ldots,m_4})^2$ subject to the constraint 
$\sum_{\alpha=0}^4 m_{\alpha}=3S$. 
$ C^{3S}_{m_0,m_1,...}$ is the CG coefficient which
describes the ground state of spin 3S in terms of the 5 original
spins S, whose expression is a product of 4 standard CG
coefficients.
The $\xi_{\alpha}$ satisfy the {\em sum rule}
$\sum_{\alpha=0}^4 \xi_{\alpha}=1$. We arrive at the following result,

\begin{eqnarray} \xi_{\alpha}(S)  = {1\over3S} {6S+1\over 8S+1}
{[(2S)!]^5\over[(8S)!]^2}\cr \times \sum_{m_1,\ldots,m_4} \;  
m_{\alpha}
{(4S-\sum_1^4 m_i)! \; [(4S+\sum_1^4 m_i)!]^2 \over \prod_1^4 (S-m_i)!
\; (S+m_i)! \; [-2S+\sum_1^4 m_i]!}  \label{dd3}
\end{eqnarray}

\noindent where if $\alpha=0$ then $m_0=3S-\sum_1^4 m_i$. It follows
that the renormalization factors for the four external spins in the
5-block are all equal $\xi_1=\xi_2=\xi_3=\xi_4\equiv \xi (S)$, while
that of the central spin $\xi_0$ is determined by the sum rule.
Amazingly enough the sum (\ref{dd3}) can be performed in a close manner
yielding,

\begin{equation} \xi (S) = {1\over3} \; {S+\case{1}{4}\over
S+\case{1}{3}} \label{dd4} \end{equation}

\noindent 
For spin $S=\case{1}{2}$ one obtains $\xi (\case{1}{2})=\case{3}{10}$.
Moreover, Eq. (\ref{dd4}) 
 correctly reproduces the classical limit $\lim_{S\rightarrow \infty}
\xi(S)=\case{1}{3}$ (recall $S=S^{\text{old}}=\case{1}{3}
S^\prime=\xi_{\text{cl}} S^\prime$). Notice also that the value for
$S=\case{1}{2}$ is already close to the classical value.

The RG-equations for the spin operators $\bf{S}_i$ $i=1,2,3,4$ allows
us to compute the renormalized Hamiltonian $H^\prime$ which turns out
to be of the same form as the original AFH Hamiltonian. In
fact, we arrive at the following RG-equations,

\begin{mathletters} \label{dd5} \begin{eqnarray} H^\prime(N,S,J) = 
 -J S (4S+1){N\over5} \cr  +
H(\case{N}{5},3 S,3\xi^2(S) J)  \label{dd5a}
\end{eqnarray} \begin{equation} N^\prime = {N\over5},\; S^\prime = 3
S, \; J^\prime = 3 \xi^2(S) J \label{dd5b} \end{equation}
\end{mathletters}

\noindent where the first contribution in Eq. (\ref{dd5a}) comes from
the energy of the blocks. As $3 \xi^2(S)<1$, the flow equation
(\ref{dd5b}) implies that the coupling constant flows to zero $J^{(n)}
\stackrel{n\rightarrow \infty}{\rightarrow} 0$  which means that the
AFH model remains {\em massless} for arbitrary value of the spin $S$.
This fact allows us to compute the density of energy $e_{\infty}(S)$
(per site) as  the following series,

\begin{mathletters} 
\label{dd6}
 \begin{eqnarray} 
e_{\infty}(S) =
-\case{1}{5} \sum_{n=0}^{\infty} \case{1}{5^n}  J^{(n)} S^{(n)} (4
\times 3^n S + 1)
 \label{dd6a} 
\end{eqnarray}
 \begin{equation}
 S^{(n+1)} = 3 S^{(n)}, \; J^{(n+1)} = 3 \xi^2(S^{(n)}) J^{(n)}
\label{dd6b}
 \end{equation} 
\end{mathletters}

Using eqs. (\ref{dd6a}) and (\ref{dd6b}) we can compute the ground state energy of
our  variational RG state for any value of the spin S. In
particular for S=1/2 we get the value $e_{\infty}$ = -0.5464. This value
has to be compared with  the ``exact'' numerical result
-0.6692, which is obtained using Green-function Monte Carlo
methods \cite{carlson}, and the spin wave value which is -0.6703. 
The difference between our result and the Green Function MC or SW
is quite big and around 0.12. To clarify the origin of this departure
we have considered the semiclassical expansion of  Eq. (\ref{dd6a})
and compare it with the standard formula of Anderson and Kubo \cite{swt},

\begin{mathletters} 
\label{dd71}
 \begin{eqnarray} 
e^{RG}_{\infty} = - 2 S ( S +  \frac{0.0223}{S} + \cdots ) 
 \label{dd71a} 
\end{eqnarray}
 \begin{equation}
 e_{\infty}^{sw} = -2 S ( S + 0.158 +
\frac{0.0062}{S} +\cdots ) 
\label{dd71b}
 \end{equation} 
\end{mathletters}

\noindent 
The important observation is that the term linear in S is absent in our 
formula (\ref{dd71a}). The reason for this is that we are using a first order
RG method for which the ground state energy follows  from the formula
$E_{\text{GS}}=\langle \Psi_0| H | \Psi_0 \rangle$, where $| \Psi_0 \rangle$
is the variational ground state  constructed by the RG method.
Now it is easy to see that taking $| \Psi_0 \rangle$ to be simply the Neel 
state one has to go to second order perturbation theory (PT) to get a linear
term in S, which turns out to be given by $S/4+1/32 + O(1/S)$.
It is clear that the ``missing energy " 0.12 is due to this peculiarity of the 
first order PT.
To remedy this one should implement the RG method with second order
PT. In 1D and for S=1/2 this can be done, obtaining for the ground state
energy density $e_{\infty} = -0.4530$ which is comparable in precision
with the spin wave result -0.4647 (recall that the exact value is -0.4431.)
The latter computation in 2D is much more involved but it is expected to
yield a result close to the spin wave result.

In order to have a better insight into the physics of the model it is
convenient to  compute the staggered magnetization  $M\equiv \langle
\case{1}{N} \sum_j (-1)^j S^z_j \rangle$.
We have been able to  obtain a
closed formula for arbitrary spin $S$ which is capable of analytical
study. To this purpose, we use the RG-equality for V.E.V.  $\langle
\psi_0| {\cal O} |\psi_0 \rangle= \langle \psi^\prime_0| {\cal
O}^\prime |\psi^\prime_0 \rangle$ for renormalized observables ${\cal
O}^\prime$ in the ground state and divide the sum in $M$ into 5-block
contributions. With the help of the renormalization spin factors we
arrive at the RG-equation for the staggered magnetization,

\begin{equation} M_N (S) = {8 \xi (S) - 1\over5} M_{N/5} (3 S)
\label{dd7} \end{equation}

\noindent The explicit knowledge of $\xi (S)$ (\ref{dd3}) allows us to
solve this  RG-equation for the staggered magnetization in the
thermodynamic limit  $N\rightarrow \infty$.  In fact, as we know by
now that the Hamiltonian renormalizes to its classical limit, we have 
$\lim_{S\rightarrow \infty} M(S)=S$. Defining $M(S) \equiv S f(S)$,
Eq. (\ref{dd7}) amounts to solving the equation $f(S)={S+1/5\over S+1/3}
f(3 S)$ subject to the boundary condition $f(\infty)=1$. Thus, we
obtain the following formula for the staggered magnetization for 
arbitrary spin,

\begin{equation} M(S) = S \; \prod_{n=0}^{\infty} {S+\case{1}{5} 3^{-n}
\over S+\case{1}{3} 3^{-n}} \label{dd8} \end{equation}

\noindent This is a nice formula in several regards. For spin
$S=\case{1}{2}$ we get $M(\case{1}{2})=0.373$ to be compared with
the most accurate Quantum Monte Carlo reslult \cite{wy} which is
0.3074 (earlier numerical results were obtained with Green function
QMC methods 
\cite{carlson} and Variational Monte Carlo plus Lanczos algorithm
\cite{hebb-rice}).
Other approximate methods
employed so far lead to values of $M(\case{1}{2})$ such as,
 e.g.,  spin
wave theory plus $1/S$-expansion to order $S^{-2}$ gives \cite{hamer,canali}
 0.3069 (earlier SW results were provided by Anderson and 
Kubo \cite{swt}), spin wave
theory  plus perturbation theory gives \cite{huse} 0.313, etc.
Our value is close to the one found\cite{parrinello}
 with pertubation theory 
around the Ising model to order 4 which is 0.371.
We can equally get values of the staggered magnetization for arbitrary
spin.
For spin S=1, our formula (\ref{dd8}) gives 0.8454 to be compared with
0.8043 using SW to order \cite{hamer} $1/S^2$  and 0.8039 obtained
by Wheihong et al. \cite{w} using series expansions.

 Another interesting feature of our formula
(\ref{dd8}) is that it allows us to make a $1/S$-expansion yielding the
result,

\begin{mathletters} 
\label{dd9}
 \begin{eqnarray} 
 M^{\text{RG}}_{\infty}(S) = 
S - 0.2 + 0.06 {1\over S} + O(1/S^2)
 \label{dd9a} 
\end{eqnarray}
 \begin{equation}
 M_{\infty}^{sw} (S)=S - 0.198 + O(1/S^2)
\label{dd9b}
 \end{equation} 
\end{mathletters}

\noindent Observe the excellent agreement between the order $S^0$ term
in both formulas. Recall that equation (\ref{dd9a}) is derived using first order
PT. We expect that a second order RG would further lower the value of 
$M(S)$, in agreement with the numerical result.

In summary we can claim that the rotating-valence-bond scenario gives
a consistent and suggestive picture of the ground state of the 2D AF-Heisenberg
model: the quantum fluctuations of the Neel state consist of rotating bonds
which appear at all scales corresponding to effective spins which renormalize
towards the classical value.


\section{Appendix: Second Order Formalism for the Standard BRG Method}

The modern fashion to include correlations between blocks in the real-space
RG is the DMRG \cite{white}. Nevertheless, there is an old way to include
those correlations which we believ has been overlooked in the past.
It amounts to include a second order contribution to the BRG. Recall that
the standard BRG in previous sections is a first order method from the point of 
view of Perturbation Theory (P.T.), i.e., the interblock Hamiltonian $H_{BB}$ is 
treated perturbatively in first order.
We can extend this treatment \cite{gyunp} to second order P.T.  in the usual fashion and
thus arrive to an effective Hamiltonian given by:

\begin{equation}
H_{eff} = P_B [H_B + H_{BB} + 
H_{BB} (1 - P_B)  {1\over E_B-H_B} (1 - P_B) H_{BB}] P_B
\label{app1}
\end{equation}

\noindent where $E_B$ is the ground state energy of the block and $P_B$ denotes
the projector onto te ground state of the block. This is nothing but the intertwiner
operator of section 2.

As a matter of illustration, we shall work out this formalism for the isotropic
AF Heisenberg model in 1D using the 3-site block BRG explained in section 8.
Moreover, we restrict to spin 1/2. Denote each block with an index $\alpha$.
Thus, the Hilbert space for each block ${\cal H}^{\alpha}$ is decomposed into
1/2, 1/2 and 3/2-spin subspaces,

\begin{equation}
{\cal H}^{(\alpha)} = {\cal H}^{(\alpha)}_0 (1/2) \oplus {\cal H}^{(\alpha)}_1 (1/2)
\oplus {\cal H}^{(\alpha)}_2 (3/2)
\label{app2}
\end{equation}

\noindent Correspondingly, we introduce 3 projectors onto those subspaces,

\begin{equation}
P^{(\alpha)}_0 + P^{(\alpha)}_1 + P^{(\alpha)}_2 = 
{\bf 1}^{(\alpha)}
\label{app3}
\end{equation}

\noindent They satisfy the following properties that we will be useful,

\begin{mathletters} 
\label{app4}
 \begin{eqnarray} 
P_B = \prod_{\alpha =1}^{N'} P_0^{(\alpha)}
 \label{app4a} 
\end{eqnarray}
  \begin{eqnarray} 
1 - P_B = 1 - \prod_{\alpha =1}^{N'} P_0^{(\alpha)} \neq 
\prod_{\alpha =1}^{N'} (1 -  P_0^{(\alpha)})
 \label{app4b} 
\end{eqnarray}
  \begin{eqnarray} 
(P^{(\alpha)}_m)^2 = P^{(\alpha)}_m, \quad 
P^{(\alpha)}_m P^{(\alpha)}_{m^\prime} = 0, \quad m \neq m' 
 \label{app4c} 
\end{eqnarray}
 \begin{eqnarray} 
P^{(\alpha)}_m P^{(\beta)}_{m^\prime} = P^{(\beta)}_{m^\prime} P^{(\alpha)}_m
 \label{app4d} 
\end{eqnarray}
\end{mathletters}
\noindent
where $N'=N/3$ and $m$ indicates the site in each block.

\noindent With the help of these properties, the second order contribution to
$H_{eff}$ in (\ref{app1}), denoted by $H_{eff}^{(2)}$, can be given the following form
containing 3 types of terms:

\begin{eqnarray}
H_{eff}^{(2)} = P_B \sum_{\alpha=1}^{N'} \sum_{m_{\alpha}\neq 0} 
{1\over e_0 - e_{m_{\alpha}}} \cr
\{ 
[(P_0^{(\alpha -1)} {\bf S}_3^{(\alpha -1)} P_0^{(\alpha -1)})\cdot
(P_0^{(\alpha)} {\bf S}_1^{(\alpha)} P_{m_{\alpha}}^{(\alpha)})] 
[(P_{m_{\alpha}}^{(\alpha)} {\bf S}_3^{(\alpha)} P_0^{(\alpha)}) \cdot
(P_0^{(\alpha +1)} {\bf S}_1^{(\alpha +1)} P_0^{(\alpha +1)}) \cr 
+
[(P_0^{(\alpha)} {\bf S}_3^{(\alpha)} P_{m_{\alpha}}^{(\alpha)})\cdot
(P_0^{(\alpha +1)} {\bf S}_1^{(\alpha +1)} P_0^{(\alpha +1)})]
[(P_0^{(\alpha -1)} {\bf S}_3^{(\alpha -1)} P_0^{(\alpha -1)}) \cdot 
(P_{m_{\alpha}}^{(\alpha)} {\bf S}_3^{(\alpha)} P_0^{(\alpha)})] \} P_B \cr 
+P_B \sum_{\alpha=1}^{N'} \sum_{(m_{\alpha},m_{\alpha+1})\neq (0,0)} 
{1\over 2 e_0 - e_{m_{\alpha}} - e_{m_{\alpha +1}}} \cr
[(P_0^{(\alpha)} {\bf S}_3^{(\alpha)} P_{m_{\alpha}}^{(\alpha)})\cdot
(P_0^{(\alpha +1)} {\bf S}_1^{(\alpha +1)} P_{m_{\alpha+1}}^{(\alpha +1)})]
[(P_{m_{\alpha}}^{(\alpha)} {\bf S}_3^{(\alpha)} P_0^{(\alpha)}) \cdot
(P_{m_{\alpha+1}}^{(\alpha +1)} {\bf S}_1^{(\alpha +1)} P_0^{(\alpha +1)})] P_B
\label{app5}
\end{eqnarray}

\noindent In order to work out this expression (\ref{app5}) towards a manageable
result, we need to perform a renormalization of the spin operators both in  both
subspaces of spin-1/2 (recall that in sect.8 we did it only for the lowest energy
spin 1/2.)

\noindent Let us introduce the following notation for the 4 states of spin 1/2:

\begin{equation}
|m,\beta \rangle \quad \text{with} \quad m = \pm 1/2, \beta = 0,1
\label{app6}
\end{equation}

\noindent Denote by ${\bf S}^\prime$ the effective spin-1/2 coming out of the
block renormalization. Then,

\begin{equation}
\langle m, \beta |{\bf S}_i|m',\beta' \rangle = \langle m|{\bf S}'|m'\rangle 
(\rho_i)_{\beta,\beta'}
\label{app7}
\end{equation}

\noindent with the $\rho$-matrices given by,

\begin{equation}
 \rho_1 = \left( \begin{array}{cc}
2/3 & -1/\sqrt{3} \\
 -1/\sqrt{3} &  0    
\end{array}              \right), \quad 
 \rho_2 = \left( \begin{array}{cc}
-1/3 & 0 \\
 0 &  1    
\end{array}              \right), \quad
\rho_3 = \left( \begin{array}{cc}
2/3 & 1/\sqrt{3} \\
 1/\sqrt{3} &  0    
\end{array}              \right)\label{app8}
\end{equation}

\noindent Thus, the spin renormalization that we were searching for is
summarized in

\begin{equation}
P_{\beta} {\bf S}_i P_{\beta'} = {\bf S}' (\rho_i)_{\beta,\beta'}
\label{app9}
\end{equation}

\noindent Namely,

\begin{mathletters} 
\label{app10}
\begin{eqnarray} 
P_0 {\bf S}_1 P_0 = P_0 {\bf S}_3 P_0 \equiv (\xi^{(0)} = {2\over 3}) {\bf S}'
\label{app10a} 
\end{eqnarray}
\begin{eqnarray} 
P_0 {\bf S}_1 P_1 = P_1 {\bf S}_1 P_0 \equiv (\xi^{(0)}_1 = {-1\over \sqrt{3}}) {\bf S}'
\label{app10b} 
\end{eqnarray}
\begin{eqnarray} 
P_0 {\bf S}_3 P_1 = P_1 {\bf S}_3 P_0 \equiv (\xi^{(0)}_3 = {1\over \sqrt{3}}) {\bf S}'
 \label{app10c} 
\end{eqnarray}
\end{mathletters}

\noindent Upon substitution of these expressions in (\ref{app5}) we are led to the
renormalization of the Hamiltonian:

\begin{equation}
H_{eff}^{(2)} = \sum_{\alpha=1}^{N'} [d^{(2)} + 
J_1^{(2)} {\bf S}_{\alpha} \cdot {\bf S}_{\alpha+1} +
J_2 {\bf S}_{\alpha} \cdot {\bf S}_{\alpha+2}]
\label{app11}
\end{equation}

\begin{equation}
H_{eff}^{(0+1)} = \sum_{\alpha=1}^{N'} [d^{(0)} + 
J_1^{(1)} {\bf S}_{\alpha} \cdot {\bf S}_{\alpha+1}]
\label{app12}
\end{equation}

\noindent with the following numerical values,

\begin{equation}
\begin{array}{lll}
d^{(0)} = -1, & J_1^{(1)} = {4\over 9} = 0.44 &  \\
d^{(2)} = -0.104861, & J_1^{(2)} = {211\over 1620} = 0.130247, &
J_2 = {10\over 243} = 0.0411523
\end{array}
\label{app13}
\end{equation}

Altogether, we end up with the following effective Hamiltonian in which the 
second order formalism employed shows up as a nearest-neighbour coupling $J_2$,

\begin{equation}
H_{eff} = \sum_{\alpha=1}^{N'} [d + 
J_1 {\bf S}_{\alpha} \cdot {\bf S}_{\alpha+1} +
J_2 {\bf S}_{\alpha} \cdot {\bf S}_{\alpha+2}]
\label{app14}
\end{equation}

\noindent with,

\begin{equation}
d = -1.104861, \quad J_1 = 0.574691, 
\quad J_2 = 0.0411523
\label{app15}
\end{equation}

\noindent We can now iterate this RG procedure as usual to obtain
the RG-flow equations for the two coupling constants $J_1$ and $J_2$:

\begin{equation}
\begin{array}{l}
J_1^{(m+1)} = a J_1^{(m)} - b J_2^{(m)} \\
J_2^{(m+1)} = c J_2^{(m)}
\end{array}
\label{app16}
\end{equation}

\[ a = 0.57491 \quad b = 0.44444 \quad c = 0.041152 \]

\noindent The fixed points of these RG-eqs. are simply,

\begin{equation}
({J_2\over J_1})_c = {1\over 2} [{a\over b} \pm \sqrt{({a\over b})^2 - 
4 ({c\over b})}]
\label{app17}
\end{equation}

\noindent Upon iteration the system flows towards the smallest fixed point,

\begin{equation}
({J_2\over J_1})_c = 0.076084
\label{app18}
\end{equation}

\noindent This happens to be an understimation of the numerical value.

Finally, we get a series expressing the ground state energy to second
order in RG,

\begin{equation}
e_{\infty} = \sum_{m=0}^{\infty} {1\over 3^{m+1}} [-\gamma_1 J_1^{(m)} +
\gamma_2 J_2^{(m)}]
\label{app19}
\end{equation}

\noindent with $\gamma_1 = -1.104861$, $\gamma_2 = 0.25$.

\noindent The above sum can be computed exactly by introducing generating
functions $J_i (x) \equiv \sum_{m=0}^{\infty} x^m J_i^{(m)}$, $i=1,2$ and 
using the RG-eqs. (\ref{app16}),

\begin{equation}
J_1 (x) = {1\over 1-ax+b c x^2}, \quad 
J_2 (x) = {cx\over 1-ax+b c x^2}
\label{app20}
\end{equation}

\noindent Thus,

\begin{equation}
e_{\infty} = -{1\over 3} {\gamma_1 - \gamma_2 {c\over 3} 
\over 1-{a\over 3} + {bc\over 9}}
\label{app21}
\end{equation}

\noindent and substituting the values of $\gamma_1$ and $\gamma_2$, we get

\begin{equation}
e_{\infty}^{(2RG)} = -0.453002
\label{app22}
\end{equation}

\noindent This is to be compared with the exact value $e_{\infty}^{(exact)}=-0.4431$
which amounts to a 2.2 \% error. This results improves even the spin wave
result $e_{\infty}^{(sw)} = -S^2 -0.36338 s -0.033011 = -0.4647$ which is a 4 \% off
the exact value. Recall that $e_{\infty}^{(1RG)} = -0.391304$ (11.6 \% .)


\bigskip
{\bf Acknowledgements}
The work presented in these notes has been done in collaboration with
Germ\'an Sierra and I have enjoyed many conversations on real-space
RG methods and related topics with him. Sections on the CBRG method are
also in collaboration with J. Rodriguez-Laguna.
I also thank R. Shankar for many comments on renormalization group during his
visit at CSIC (Madrid).
I acknowledge many useful discussions with the lecturers at the El Escorial
Summer School, specially Steve White, T. Nishino, A. Sandvik.
  
\noindent This work has been partially supported in part by CICYT under 
 contract AEN93-0776.


%
%
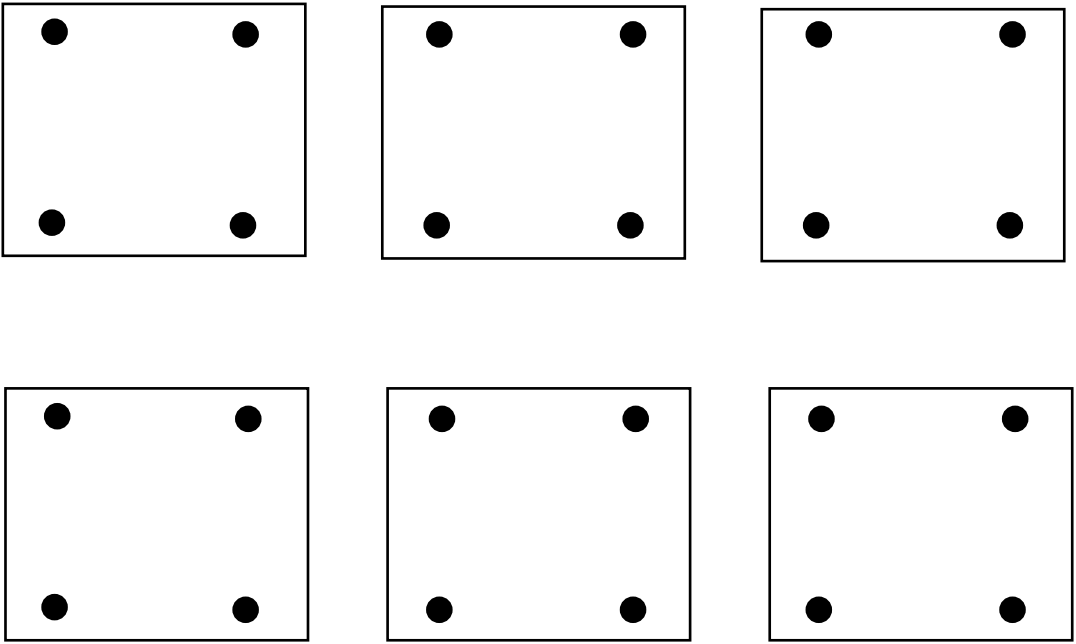
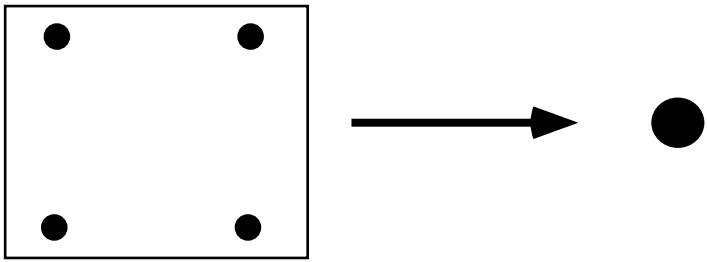
\begin{figure}
\caption{a) Block decomposition of a square lattice into 4-site blocks.
b)Squematic truncation of states in the BRG method associated to the 
previous lattice decomposition.
\label{f1}
}
\end{figure}

%
%
\begin{figure}
\caption{
a) Ground state $\psi _0$ and first excited
state $\psi _1$ for the Hamiltonian  $H_{Fixed}$ with
fixed BC's.
b) Ground state $\psi _0$ and first
excited state $\psi _1$ for the Hamiltonian  $H_{Free}$
with free BC's.
\label{f2}
}
\end{figure}

%
%
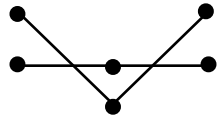
\begin{figure}
\caption{ Building blocks of the 3-site BRG for the tight-binding
model in 1D with free BC's. a) Ground state, b) First excited estate,
c) Second excited state.
\label{f3}
}
\end{figure}

%
%
\begin{figure}
\caption{Pictorical decomposition of a given Hamiltonian 
$H$ into
uncorrelated  $A$-operators, correlation $B_L$- $B_R$-operators 
and interaction $C$-operators according to the CBRG method. $B_1$
is a superblock made up of two $L_1$ and $R_1$ blocks.
\label{f4}
}
\end{figure}

%
%
\begin{figure}
\caption{The $n^2/N^2$-law 
for the first
5 excited states of the 1D Tight-Binding  Model for a chain of
$N=12 \times 2^m$ sites with Free-Free BC's. This is a 
$\ln E_n$-$\ln m$ plot.
\label{f5}
}
\end{figure}

%
%
\begin{figure}
\caption{The  wave function reconstruction for the first
5 excited states of the 1D Tight-Binding  Model for a chain of
$N=12 \times 2^6=768$ sites with Free-Free BC's. We have scaled 
up the exact results by a factor of 1.23 for clarity.
\label{f6}
}
\end{figure}

%
%
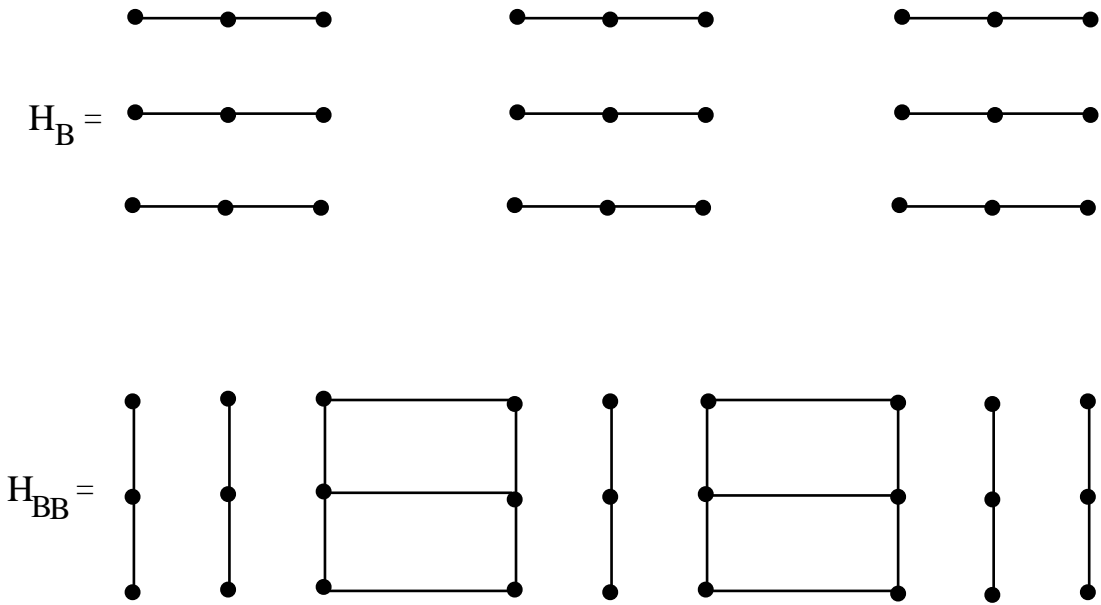
\begin{figure}
\caption{ Block decomposition associated to the standard BRG method
applied to the uniform 3-leg Heisenberg ladder.
\label{f7}
}
\end{figure}

%
%
\begin{figure}
\caption{a) The antiferromagnetic 5-block state.
b) Formation of a rotating-valence-bond state upon applying the $H_{xy}$ part 
of the Hamiltonian to the AF 5-block state.
\label{f8}
}
\end{figure}
%
%
\begin{figure}
\caption{Artistic tesselation of the square lattice with 5-site blocks.
\label{f9}
}
\end{figure}
%
%
\begin{figure}
\caption{The two-dimensional square
lattice tesselated by the 5-block.
Dahed lines are nearest-neighbours in the 
renormalized lattice.
\label{f10}
}
\end{figure}

\newpage



\begin{table}[p] 
\centering 
\begin{tabular}{|c|c|c|c|} 
\hline \hline
  \mbox{Energies} & \mbox{ Exact}    
&  \mbox{CBRG} & \mbox{DMRG} \\  \hline \hline
 $E_0$& $0 $  &
$1.1340\times 10^{-14}$ & $1.0\times 10^{-6}$ \\ \hline 
$E_1$ & $1.6733 \times 10^{-5}$  & $1.9752\times
10^{-5}$  & $1.6733 \times 10^{-5}$ \\  \hline
 $E_2$ & $6.6932 \times 10^{-5}$  &
$7.6552\times 10^{-5}$ & $6.6932 \times 10^{-5}$\\ \hline
 $E_3$& $1.5060 \times 10^{-4}$  &
$1.8041\times 10^{-5}$  & $1.5060 \times 10^{-4}$ \\ \hline
 $E_4$& $2.6772 \times 10^{-4}$  &
$2.9681\times 10^{-4}$ & $2.6772 \times 10^{-4}$ \\ \hline 
 $E_5$& $4.1831 \times 10^{-4}$   &
$5.1078\times 10^{-4}$ & $4.1831 \times 10^{-4}$    \\ \hline \hline 
\end{tabular}
\caption{Exact and CBRG Values of  Low Lying 
States for the 1D Tight-Binding  Model for a chain of
$N=12 \times 2^6=768$ sites with Free-Free BC's. 
DMRG values are also given.}  \end{table}

\begin{table}[p] 
\centering 
\begin{tabular}{|c|c|c|c|c|} 
\hline \hline
  \mbox{m} & \mbox{N=12 $2^m$}    & $E_1^{(exact)}(N)$ &
$E_1^{(CBRG)}(N)$ & $E_1^{(DMRG)}(N)$ \\  \hline \hline
 $0$& $12$ & $6.8148\times 10^{-2}$& $6.8148\times 10^{-2}$ & $6.8148\times 10^{-2}$
\\ \hline 
$1$ & $24$ & $1.7110 \times 10^{-2}$
&$1.7375\times 10^{-2}$ & $1.7110 \times 10^{-2}$
 \\  \hline
 $2$ & $48$& $4.2826 \times 10^{-3}$  &  
$4.4694\times 10^{-3}$ & $4.2826 \times 10^{-3}$ \\ \hline 
$3$ &$96$ & $1.0708\times 10^{-3} $ &
$1.1515\times 10^{-3}$  & $1.0708\times 10^{-3} $ \\  \hline 
$4$ & $192$ & $2.6772 \times 10^{-4}$ & $2.9681 \times
10^{-4}$ & $2.6772 \times 10^{-4}$  \\  \hline
$5$ & $384$ & $6.6932 \times 10^{-5}$ 
& $7.6552 \times 10^{-5}$  & $6.6932 \times 10^{-5}$   \\  \hline
$6$ & $768$ & $1.6733 \times 10^{-5}$ &
 $1.9752 \times 10^{-5}$   & $1.6733 \times 10^{-5}$ \\  \hline
  & $\gg 1$ & $\pi^2/N^2$ & $9.8080/N^2$ & $9.8696/N^2$ \\ \hline \hline
\end{tabular} \caption{Exact  and new CBRG values
of the first excited state
 for the 1D Tight-Binding  Model with Free-Free BC's.
DMRG values are also given.}  \end{table}



\begin{table}[p] 
\centering 
\begin{tabular}{|c|c|c|c|} 
\hline \hline
  \mbox{Energies} & \mbox{ Exact}    & \mbox{ Standard BRG}
&  \mbox{CBRG} \\  \hline \hline
 $E_0$& $1.7754\times 10^{-5}$ & $1.5771\times 10^{-2}$  &
$1.8409\times 10^{-5}$  \\ \hline 
$E_1$ & $1.5043\times 10^{-4}$ &
 $4.2679\times 10^{-2} $ & $1.6655\times
10^{-4}$  \\  \hline
 $E_2$ & $4.1761 \times 10^{-4}$ & $4.2794\times 10^{-2}$ &
$4.6408\times 10^{-4}$ \\ \hline
 $E_3$& $8.1831\times 10^{-4}$ & $4.3053\times 10^{-2}$  &
$9.1450\times 10^{-4}$  \\ \hline
 $E_4$& $1.3520\times 10^{-3}$ & $4.3173\times 10^{-2}$  &
$1.5179\times 10^{-3}$  \\ \hline 
 $E_5$& $2.0196\times 10^{-3}$ & $4.4288\times 10^{-2}$  &
$2.2852\times 10^{-3}$   \\ \hline \hline 
\end{tabular}
\caption{Exact, Standard RG and CBRG Values of  Low Lying 
States for the 1D Tight-Binding  Model for a chain of
$N=12 \times 2^5=384$ sites with Free-Fixed BC's.}  \end{table}

\newpage



\begin{table}
\centering 
\begin{tabular}{|c|c|c|c|c|} 
\hline \hline
  \mbox{Energies} & \mbox{ Exact}    & \mbox{ Standard BRG}
&  \mbox{CBRG} &  \mbox{DMRG} \\  \hline \hline
 $E_0$& $6.6585\times 10^{-5}$ & $5.8116\times 10^{-2}$  &
$7.0843\times 10^{-5}$ & $6.7\times 10^{-5}$ \\ \hline 
$E_1$ & $2.6633\times 10^{-4}$ &
 $5.8155\times 10^{-2} $ & $2.9403\times
10^{-4}$ & $2.66\times 10^{-4}$ \\  \hline
 $E_2$ & $5.9924\times 10^{-4}$ & $5.8268\times 10^{-2}$ &
$6.3690\times 10^{-4}$ & $5.99\times 10^{-4}$\\ \hline
 $E_3$& $1.0653\times 10^{-3}$ & $5.8470\times 10^{-2}$  &
$1.2289\times 10^{-3}$ & $1.065\times 10^{-3}$ \\ \hline
 $E_4$& $1.6644\times 10^{-3}$ & $5.8717\times 10^{-2}$  &
$1.7707\times 10^{-3}$  & $1.664\times 10^{-3}$\\ \hline 
 $E_5$& $2.3966\times 10^{-3}$ & $5.9106\times 10^{-2}$  &
$2.7311\times 10^{-3}$ & $2.397\times 10^{-3}$  \\ \hline \hline 
\end{tabular}
\caption{Exact, Standard RG and CBRG Values of  Low Lying 
States for the 1D Tight-Binding  Model for a chain of
$N=12 \times 2^5=384$ sites with Fixed-Fixed BC's.
 DMRG values are also given.}  \end{table}



\begin{table}
\centering 
\begin{tabular}{|c|c|c|} 
\hline \hline
  \mbox{Energies} & \mbox{ Exact}    
&  \mbox{CBRG} \\  \hline \hline
 $E_0$& $0$   &
$9.6114\times 10^{-35}$  \\ \hline 
$E_1$ & $1.5056\times 10^{-4}$  & $1.9390\times
10^{-4}$  \\  \hline
 $E_2$ & $1.5056\times 10^{-4}$  &
$1.9390\times 10^{-4}$ \\  \hline
 $E_3$& $3.0012\times 10^{-4}$  &
$3.8781\times 10^{-4}$  \\ \hline \hline 
\end{tabular}
\caption{Exact and CBRG Values of  Low Lying 
States for the 2D Tight-Binding  Model for a lattice of
$N=4\times 4 \times 4^6=65536$ sites with Free BC's.}
  \end{table}


\begin{table}
\caption{Fixed Points of the Anisotropic AF-Heisenberg Model}
\vspace{2pt}
\begin{tabular}{ccccl}
\hline
\multicolumn{1}{c}{\rule{0pt}{12pt}%
       $\Delta$}&\multicolumn{1}{c}{%
       $0$}&\multicolumn{1}{c}{%
       $1$}&\multicolumn{2}{c}{%
       $\infty$}\\[2pt]
\hline\rule{0pt}{12pt}
$R^{BRG}_{\infty}$& $-0.2828$ & $-0.3913$  & $-\frac{1}{4} \Delta $  & \\
$e^{exact}_{\infty }$ & $-0.3183$ & $-0.4431$ & $-\frac{1}{4} \Delta $ &  \\
$\frac{e^B - e^{exact}}{e^{exact}}\times 100$ & $11\%$ & $12\%$ & 0& \\[2pt]
\hline
\end{tabular}
\end{table}


\begin{references}

\bibitem{white}  S.R. White, 
 {\em Phys. Rev.  Lett.} {\bf 69}, 2863 (1992);
 {\em Phys. Rev.  B} {\bf 48}, 10345 (1993).

\bibitem{whitenote} S.R. White,
Proceedings of El Escorial Summer School 1996, and references therein.

\bibitem{white2D} S.R. White, Preprint cond-mat/9604129.

\bibitem{nishinonote} T. Nishino, ``Density Matrix and Renormalization
for Classical Lattice Models", proceedings in this volume.

\bibitem{pereznote} J. P\'erez-Conde,
Proceedings of El Escorial Summer School 1996, and references therein.


\bibitem{wilson} K.R. Wilson,  {\em  Rev. Mod. Phys.} 
{\bf 47}, 773 (1975).

\bibitem{anderson1} P.W. Anderson, 
{\em J. Phys. C} {\bf 3}, 2436 (1970).


\bibitem {hirsch} Hirsch, J. {\em 1983,  Phys. Rev. B {\bf
28}, 4059}, {\em 1985,  Phys. Rev. B {\bf 31}, 4403};
Hirsch, J. and Lin, H.Q., {\em 1988,  Phys. Rev. B {\bf
37}, 5070}.

\bibitem{drell} S.D. Drell, M. Weinstein, S. Yankielowicz, 
 {\em Phys. Rev.  D} {\bf 16}, 1769 (1977).

\bibitem{jullien} R. Jullien, P. Pfeuty, J.N. Fields, S.
Doniach,
 {\em Phys. Rev.  B} {\bf 18}, 3568 (1978).


\bibitem {hirschII} Hirsch, J. {\em 1980,  Phys. Rev. B
{\bf 22}, 5259}.


\bibitem{dm-germanyo} M.A. Mart\'{\i}n-Delgado and G. Sierra, 
``Analytic Formulations of the Density Matrix Renormalization Group", 
Int. J. Mod. Phys. {\bf A11} 3145, (1996).


\bibitem{jullienlibro}  ``Real-Space Renormalization", editors Burkhardt, T.W. and van
Leeuwen, J.M.J., series topics in Current Physics {\bf 30},
Springer-Verlag 1982.

\bibitem{jaitisi} J. Gonz\'alez, M.A. Mart\'{\i}n-Delgado,  
 G. Sierra, A.H. Vozmediano, \\
{\em Quantum Electron Liquids and
High-$T_c$ Superconductivity}, \\
Lecture Notes in Physics, Monographs
vol. {\bf 38}, Springer-Verlag 1995.



\bibitem {q-germanyo} M.A. Mart\'{\i}n-Delgado and G. Sierra, 
``Real Space Renormalization Group Methods and
Quantum Groups". Phys. Rev. Lett. {\bf 76},1146 (1996).

\bibitem {qbis-germanyo} M.A. Mart\'{\i}n-Delgado and G. Sierra,
 in ``From Field Theory to Quantum Groups". World Scientific
Publishers 1996.



\bibitem{white-noack} S.R. White, R.M. Noack,
 {\em Phys. Rev.  Lett. }{\bf 68}, 3487 (1992).

\bibitem {wilson-unp} Wilson, K.G.,
{\em 1986,  unpublished informal talk.}

\bibitem{kl} V. Karimipour and A. Langari, 
``A Modified Quantum  Renormalization Group for the XXZ Spin Chain", 
preprint 1996, private communication.

\bibitem {rabin}J.M. Rabin,
 {\em Phys. Rev.  B} {\bf 21}, 2027 (1980).


\bibitem {bc-germanyo} M.A. Mart\'{\i}n-Delgado and G. Sierra, 
``The Role of Boundary Conditions in the 
Real-Space Renormalization Group".
 Phys. Lett. {\bf B364} 41, (1995).

\bibitem {cbrg} M.A. Mart\'{\i}n-Delgado, J. Rodriguez-Laguna
 and G. Sierra, 
``The Correlated Block Renormalization Group".
 Nucl. Phys. {\bf B473} 685, (1996).

\bibitem {fpacheco}  A. Fern\'andez-Pacheco, {\em Phys. Rev.  D}
{\bf  19}, 3173 (1979).


\bibitem{nishino1}T.~Nishino,
``Density Matrix Renormalization Group Method for 2D Classical Models",
 J.~Phys.~Soc.~Jpn.~{\bf 64} (1995) 3958-3961.




\bibitem{nishino} T. Nishino and K. Okunishi,
 ``Corner Transfer Matrix Renormalization Group Method",
J.~Phys.~Soc.~Jpn.~{\bf 65} (1996) 891-894. Cond-mat/9507087.


\bibitem{ostlund-rommer} S. Ostlund  and S. Rommer,
 ``Thermodynamic limit of the density matrix renormalization for 
the spin-1 Hesisenberg chain". Phys. Rev. Lett. {\bf 75}, 3537 (1996).

\bibitem{nishino-okunishi} T. Nishino  and K. Okunishi,
 ``Product Wave Function Renormalization Group". 
J.~Phys.~Soc.~Jpn.~{\bf 64} (1995) 4084-4087. Cond-mat/9510004.

\bibitem {lieb-s-m} Lieb, E., Schultz, T. and Mattis, D.
{\em 1961,  Ann. Phys. {\bf 16}, 407}

\bibitem {orbach} Orbach, R.,
{\em 1958,  Phys. Rev.  {\bf 112}, 309}

\bibitem {gyunp} M.A. Mart\'{\i}n-Delgado and G. Sierra,
unpublished work 1996.

\bibitem{prb86}
H.~J. Schulz, Phys. Rev. B {\bf 34},  6372  (1986).

\bibitem{schulznote} H.J. Schulz,
Proceedings of El Escorial Summer School 1996, and references therein.

\bibitem{sierranote} G. Sierra,
Proceedings of El Escorial Summer School 1996, and references therein.

\bibitem {rotating}  G. Sierra and M.A. Mart\'{\i}n-Delgado,
``Real Space Renormalization Group Approach to the
2D Antiferromagnetic Heisenberg Model", preprint (1996).

\bibitem{manousakis} E. Manousakis, Rev. Mod. Phys. {\bf 63}, 1 (1991). 

\bibitem{bethe} H.A. Bethe, Z. Phys. {\bf 71}, 205 (1931).

\bibitem{anderson} P.W. Anderson, Science {\bf 235}, 1196 (1987).

\bibitem{kivelson}  S.A. Kivelson, D.S. Rokhsar and J.P. Sethna,
 {\em Phys. Rev.  B} {\bf 35}, 8865 (1987).


\bibitem{liang-doucot-anderson} S. Liang, B. Doucot and P.W. Anderson,
{\em Phys. Rev.  Lett.} {\bf 61}, 365 (1988).


\bibitem{carlson}  J. Carlson,
 {\em Phys. Rev.  B} {\bf 40}, 846 (1989).

\bibitem{hebb-rice}  Hebb and T. M. Rice,
 {\em Z. Phys.  B} {\bf 90}, 73 (1993).

\bibitem{white-noack-scalapino} S. White, R. Noack and D. Scalapino,
{\em Phys. Rev.  Lett.} {\bf 73}, 886 (1994).

\bibitem{dagotto-rice} E. Dagotto and T.M. Rice,
{\em Science} {\bf 271}, 618 (1996).

\bibitem{anderson-fazekas} P. Fazekas and P.W. Anderson,
{\em Phil. Mag.} {\bf 30}, 432 (1974)

\bibitem{CHN} S. Chakravarty, B.I. Halperin and D.R. Nelson,
{\em Phys. Rev.  Lett.} {\bf 60}, 1057 (1988).

\bibitem{wy} Wise and Ying, 
 {\em Z. Phys.  B} {\bf 93}, 147 (1994).

\bibitem{hamer} C.J. Hamer, Z. Weihong and P. Arndt,
{\em Phys. Rev.  B} {\bf 46}, 6276 (1992).


\bibitem{canali} C.M. Canali and M. Vallin,
{\em Phys. Rev.  B} {\bf 48}, 3264 (1992).

\bibitem{swt}  P.W. Anderson,
 {\em Phys. Rev.  B} {\bf 6}, 694 (1952); R. Kubo, {\em Phys. Rev.}
{\bf 87}, 568 (1952).

\bibitem{huse}  D. Huse,
 {\em Phys. Rev.  B} {\bf 37}, 2380 (1988).


\bibitem{parrinello} M. Parrinello and T. Arai,
 {\em Phys. Rev.  B} {\bf 10}, 265 (1974).


\bibitem{w}  Z. Weihong, J. Oitmaa and C.J. Hamer,
{\em Phys. Rev.  B} {\bf 43}, 8321 (1991).




\end{references}
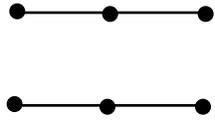
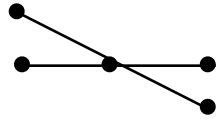
\end{document}